\documentclass[sn-mathphys-num]{sn-jnl}

\usepackage{graphicx}%
\usepackage{multirow}%
\usepackage{amsmath,amssymb,amsfonts}%
\usepackage{amsthm}%
\usepackage{mathrsfs}%
\usepackage[title]{appendix}%
\usepackage{xcolor}%
\usepackage{textcomp}%
\usepackage{manyfoot}%
\usepackage{booktabs}%
\usepackage{algorithm}%
\usepackage{algorithmicx}%
\usepackage{algpseudocode}%
\usepackage{listings}%
\usepackage{xspace}  

\begin{document}

\newcommand{\system}{ARTHUR\xspace} 
\newcommand{\systemLong}{\textbf{A}utho\textbf{r}ing \textbf{T}ool for \textbf{Hu}man--\textbf{R}obot collaboration scenarios\xspace}

\newcommand\todo[1]{\textcolor{blue}{TODO: #1}}
\newcommand\draft[1]{\textcolor{gray}{\noindent DRAFT: #1}}
\newcommand\highlight[1]{\textcolor{orange}{#1}}
\newcommand\rev[1]{#1}

\newcommand{\feedback}{\textit{feedback}\xspace} 
\newcommand{\Feedback}{\textit{Feedback}\xspace}

\newcommand{\actions}{\textit{actions}\xspace}
\newcommand{\Actions}{\textit{Actions}\xspace} 
\newcommand{\action}{\textit{action}\xspace} 

\newcommand{\conditions}{\textit{conditions}\xspace}
\newcommand{\Conditions}{\textit{Conditions}\xspace}
\newcommand{\condition}{\textit{condition}\xspace}

\newcommand{\phaseOne}{\textit{configuration}\xspace}
\newcommand{\phaseTwo}{\textit{refinement}\xspace}
\newcommand{\phaseThree}{\textit{operation}\xspace}

\newcommand{\nVisualisations}{20\xspace}
\newcommand{\nActions}{10\xspace}
\newcommand{\nConditions}{18\xspace}
\newcommand{\nProperties}{11\xspace}

\newcommand{\lego}{a world-leading toy manufacturer}

\newcommand{\repo}{\url{https://gitlab.au.dk/arthur}\xspace}
\newcommand{\repoVal}{2}
\newcommand{\video}{\url{https://www.youtube.com/playlist?list=PLhjFAueqW0cuRR-ZruATwfFXfasdi5Gli}\xspace}
\newcommand{\videoVal}{3}

\title[Authoring Human-Robot Collaboration Processes]{ARTHUR: Authoring Human-Robot Collaboration Processes with Augmented Reality using Hybrid User Interfaces}


\author*[1]{\fnm{Rasmus} \sur{Lunding}}\email{rsl@cs.au.dk}

\author[2]{\fnm{Sebastian} \sur{Hubenschmid}}\email{Sebastian.Hubenschmid@uni-konstanz.de}

\author[1,2]{\fnm{Tiare} \sur{Feuchtner}}\email{tiare.feuchtner@uni-konstanz.de}

\author[1]{\fnm{Kaj} \sur{Grønbæk}}\email{kgronbak@cs.au.dk}

\affil*[1]{\orgdiv{Department of Computer Science}, \orgname{Aarhus University}, \orgaddress{
\country{Denmark}}}

\affil[2]{\orgdiv{Department of Computer and Information Science}, \orgname{University of Konstanz}, \orgaddress{
\country{Germany}}}



\abstract{

While augmented reality shows promise for supporting human-robot collaboration, creating such interactive systems still poses great challenges.
Addressing this, we introduce \system, an open-source authoring tool for augmented reality-supported human-robot collaboration. 
\system supports \nVisualisations~types of multi-modal \feedback to convey robot, task, and system state, \nActions~\actions that enable the user to control the robot and system, and \nConditions~\conditions for \feedback customization and triggering of \actions.
By combining these elements, users can create interaction spaces, controls, and information visualizations in augmented reality for collaboration with robot arms.
With \system, we propose to combine desktop interfaces and touchscreen devices for effective authoring, with head-mounted displays for testing and in-situ refinements. 
To demonstrate the general applicability of \system for human-robot collaboration scenarios, we replicate representative examples from prior work. Further, in an evaluation with five participants, we reflect on the usefulness of our hybrid user interface approach and the provided functionality, highlighting directions for future work.
}




\maketitle

\begin{figure}
    	\centering
	\includegraphics[width=\linewidth]{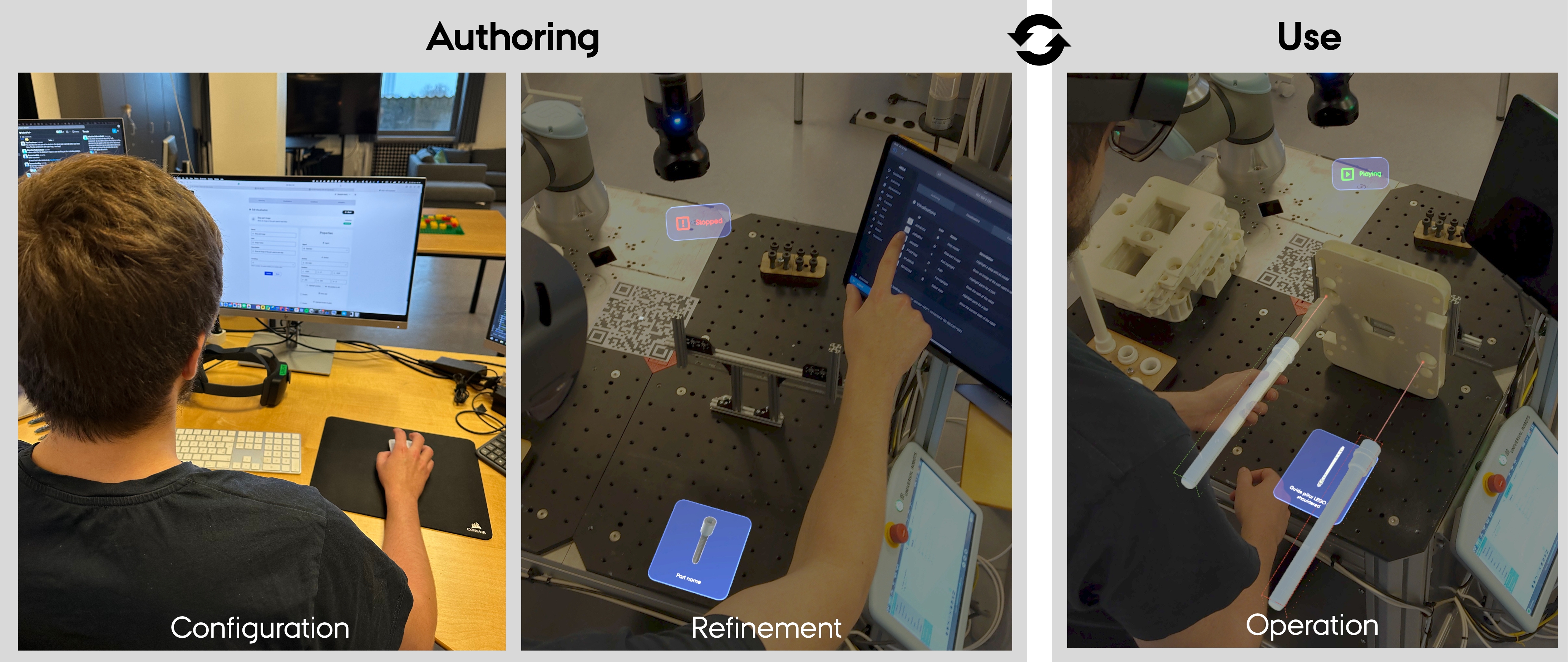}
	\caption{\system is an authoring tool for augmented reality-supported human-robot collaboration. It supports (1) creating the initial system configuration in the web interface (PC) (left), (2) refining the setup in-situ using a hybrid of web interface (tablet) and AR interface (HMD) (center), and (3) testing and using the authored system on the AR interface (HMD) (right).}
	\label{fig:teaser}
\end{figure}

\section{Introduction}
Recent research in human-robot interaction (HRI) has demonstrated the benefits of collaborative robots (cobots) for assisting with assembly tasks~\cite{lunding2023ar,malik2022drive}, as they enable close collaboration between humans and robots without extensive safeguards~\cite{cheon2022robots}. 
However, despite great interest and desire to adopt these new technologies, the manufacturing industry struggles to integrate collaborative robots into their processes for various reasons, such as lacking knowledge and safety concerns. From our industry partners in the MADE FAST project\footnote{MADE FAST: \url{https://www.made.dk/en/made-fast/}} we have observed that when
cobots are introduced they are rarely collaborated with, as the human worker only steps in to troubleshoot or takes over parts of the task within a separate workspace in turn-taking fashion.
To overcome such limitations and make ``genuine'' human-robot collaboration (HRC) possible, the operator must be made aware of the ongoing and planned robot procedures and be able to coordinate these. Information presented in augmented reality (AR) can be beneficial for conveying the robot's intent~\cite{pascher2023how,suzuki2022augmented}, visualizing safety information~\cite{ganesan2018better,hietanen2020ar}, and highlighting the procedures or tasks which the user has to do~\cite{dimitropoulos2021operator,lunding2023ar}. 
Researchers have investigated the potential of AR for supporting collaboration~\cite{ganesan2018better,rosen2019communicating}, thereby exploring the suitability of different devices~\cite{hietanen2020ar,rosen2019communicating} and individual visualization types~\cite{arevaloarboleda2021assisting,cogurcu2023augmented,gruenefeld2020mind}, or proposing systems for solving specific assembly tasks~\cite{andronas2021multi,ganesan2018better}. 
 
Notably, only few researchers have so far explored the usefulness of visualization combinations, noting drawbacks such as redundancy (e.g., convey planned robot movement through a path visualization and holograms of task-relevant objects)~\cite{lunding2023ar,rosenholtz2007measuring} and the need for dynamic visualizations that show up only when relevant~\cite{lunding2024robovisar}. 
We argue that, in addition to investigating static visualizations in isolation (e.g., \cite{cleaver2021dynamic,cogurcu2023augmented,rosen2019communicating}), we must evaluate visualizations within the context of the overall system and throughout representative workflows, to better understand the users' needs.
However, the process of designing, testing, debugging, and refining AR content is often tedious and challenging: AR setups for HRI are typically authored on a desktop system, requiring accurate simulation of the real-world environment in which the virtual content should be anchored. Even then, there is often a notable discrepancy between the simulation of AR content and its situated visualization upon deployment on the AR head-mounted display (HMD). 

To break out of the sequential design-deploy-refine process that requires time-consuming alternation between the personal computer (PC) and the HMD, we have recently demonstrated the potential of authoring visualizations directly in AR (i.e., \textit{in-situ})~\cite{lunding2024robovisar}. Users can thereby immediately test out variations of visualizations and their appearance properties, without needing to recompile and deploy the application. 
On the other hand, in-situ authoring in AR can limit the user's effectiveness compared to desktop-based interaction: mid-air text entry on a virtual keyboard is inferior to physical typing~\cite{grubert2023text}, and manipulation of traditional interface elements (e.g., buttons and lists) can be strenuous and difficult due to imprecise mid-air interaction~\cite{chan2010touching}. Recent works have therefore argued for combining multiple devices as complementary interfaces~\cite{elmqvist2023data,zagermann2022complementary} (e.g., AR HMDs and mobile touchscreen devices) to leverage the advantages of each technology. In this paper, we propose adopting this approach for HRC.

Addressing the aforementioned challenges, we present \system, 
an AR-based \systemLong that supports the creative workflow through a hybrid user interface (i.e., across a desktop, tablet, and HMD).
We expand upon prior work by facilitating in-situ authoring not only of visualizations (\feedback) but also of user \actions and \conditions, thus creating a holistic AR environment for designing HRC processes.
Besides enabling the general setup and in-situ authoring, our proposed system fluidly facilitates switching to the operation phase, allowing users to instantaneously try out their authored workflows on one or more robots. 
We demonstrate the potential of \system by replicating a variety of scenarios from prior work. Further, we evaluated the usefulness of the supported features and the suitability of our hybrid user interface approach in a usage evaluation with experts.

In summary, we contribute (1)~an AR authoring tool for HRC based on three types of design components (\feedback, \actions, \conditions) that supports (2)~a flexible authoring workflow through a hybrid user interface. Further, we highlight opportunities and challenges in authoring HRC systems with a hybrid user interface approach through (3)~ demonstrations and an expert evaluation using \system. The source code\footnote{ARTHUR project repository: \repo} and supplementary videos\footnote{ARTHUR video playlist: \video} are available online.
\section{Related work}
    We review existing approaches for \textit{AR authoring for HRI}, discuss \textit{design elements} for AR-supported HRI, and review the concept of \textit{hybrid user interfaces}.

    \subsection{AR Authoring for HRI}
        Recent surveys~\cite{ens2019revisiting, sereno2020collaborative, fidalgo2023survey, ratcliffe2021extended, marques2022conceptual, marques2024towards} show that collaborative mixed reality systems (e.g., for HRC) are now mature enough to \textit{``focus deeply on the nuances of supporting collaboration''}~\cite{ens2019revisiting}, such as the use of heterogeneous hardware~\cite{sereno2020collaborative}, moving between reality and AR during collaborative tasks~\cite{fidalgo2023survey}, or the potential for in-depth data collection~\cite{ratcliffe2021extended}.
        In the context of authoring for HRI, AR environments have primarily been explored to assist non-technical users in defining the robots' behavior. Noteworthy authoring tools include GhostAR~\cite{cao2019ghostar}, V.Ra~\cite{cao2019v.ra}, KineticAR~\cite{fuste2020kinetic}, and PRogramAR~\cite{ikeda2024programar}. However, these are concerned with robot programming and aspects related to that, but not the authoring of the AR content that supports HRI itself. 
        For authoring AR content, Microsoft Dynamics 365 Guides~\cite{microsoft} is an interesting commercial tool that allows users to author assembly instructions. However, it does not incorporate robots by interfacing with respective controllers and integrating their sensor data. A third relevant category are tools for data visualization or debugging, such as ARViz~\cite{hoang2022arviz,ikeda2022advancing}. These often provide visualization functionality that is customized towards developers and  do not necessarily address usability challenges, e.g., visual clutter and supporting user input. In this context, recent works~\cite{lee2023design, martins2022augmented} highlight how such situated visualizations can be integrated into a working environment to support decision-making.
        
        We partially addressed these shortcomings in our prior work on RoboVisAR~\cite{lunding2024robovisar}: an AR authoring tool that enables users to create situated visualizations of robot data (e.g., status and movement path). It employs a timeline-based approach where a recording is made by executing the robot program before the authoring process begins, as was also proposed in prior work~\cite{leiva2021rapido}. AR content is then designed based on this recording before deployment for live execution, while interaction and feedback are handled entirely through an AR-HMD. 
        In our view, RoboVisAR still has three key limitations for being used in a broader context, e.g. collaborative assembly, as it is: (1)~restricted to `robot visualizations', thus excluding visualization of additional/external content, such as assembly instructions; (2)~user input is not possible while the system is running (i.e., unidirectional instead of bidirectional information flow from robot to operator); and (3)~it suffers from challenges of mid-air interaction, in particular when manipulating 2D user interface elements during authoring~\cite{lunding2024robovisar, lunding2023ar}. We aim to address these limitations with \system, which supports assembly instructions, input from the user to the system, and a hybrid authoring interface to limit virtual menu interaction.

    \subsection{Design Components for AR-supported HRI}
        Recent reviews reveal that a broad range of visual design components exist not only within AR-supported human-centered collaboration~\cite{ghamandi2023what, marques2022conceptual}, but also AR-supported HRI~\cite{suzuki2022augmented, walker2023virtual}. Suzuki et al.~\cite{suzuki2022augmented} identify three groups of design components: \textit{UIs and widget}, \textit{spatial references and visualizations}, and \textit{embedded visual effects}, each consisting of further subcategories. Some of these components are highly relevant for collaborative assembly: points and locations~\cite{chan2020augmented}, paths and trajectories~\cite{andronas2021multi,lunding2023ar}, areas and boundaries~\cite{ganesan2018better,hietanen2020ar}, information panels~\cite{lunding2023ar}, and labels and annotations~\cite{andersen2016projecting}. 
        The virtual design element taxonomy by Walker et al.~\cite{walker2023virtual} provides a further classification that can be used for HRI in mixed reality and also includes task-related design elements. 
        Prior work by Li et al.~\cite{li2019research} breaks down visualizing relevant information regarding tasks, parts, tools, and processes by describing how abstract representations (e.g., arrows) can be integrated with 3D models.
        These review papers informed the \feedback types implemented by \system. 
                
        Besides feedback for the user, Suzuki et al.~\cite{suzuki2022augmented} categorize different levels of interactivity, ranging from \textit{only output}, to \textit{implicit}, \textit{explicit and indirect}, and \textit{explicit and direct}. With \system we aim is to support all these levels, such that the users can choose whatever their scenario requires. Suzuki et al.~\cite{suzuki2022augmented} further categorize interaction modalities and techniques, gaze, gesture, and proximity, most of which are supported in \system.
        
    \subsection{Hybrid User Interfaces}
         Hybrid user interfaces employ cross-device interaction~\cite{brudy2019crossdevice} to combine \textit{``heterogeneous display and interaction device technologies''}~\cite{feiner1991hybrid}, such as using AR HMDs simultaneously with smartphones, tablets, or desktop systems. The potential of AR in HRI makes this combination especially compelling, as commonly-used devices (e.g., tablets) can be seamlessly extended with superimposed content (e.g., \cite{hubenschmid2023smartphone,langner2021marvis, reipschlager2021personal}). In addition, hybrid user interfaces have been found to facilitate better performance for two-dimensional input such as text entry~\cite{grubert2023text} and navigation~\cite{buschel2019investigating}, likely due to high familiarity~\cite{butscher2018clusters,hubenschmid2021stream}, high input accuracy, and the availability of haptic feedback~\cite{knierim2021smartphone}. In the context of authoring and deploying AR systems, recent work has explored the asynchronous use of hybrid user interfaces ~\cite{hubenschmid2021asynchronous}, involving switching between an authoring environment on one device (e.g., a desktop computer) and content inspection on another (e.g., AR-HMD): For example, Hubenschmid et al.~\cite{hubenschmid2022relive} combined a familiar 2D desktop interface for visual analytics with immersive virtual reality, to allow traditional data visualization on a 2D screen, as well as immersive visualizations in-situ. The user can thereby flexibly switch to the appropriate interface (i.e., 2D ex-situ or 3D in-situ), as the tasks demand. This approach may also prove beneficial for authoring HRC processes, allowing users to switch from a familiar setup on a desktop for the initial programming to an in-situ approach for inspection and fine adjustments of the content.

         However, prior studies on such distributed systems also highlight disadvantages, such as increased cognitive load due to repeated attention switches~\cite{kim2009simulated, rogers1995costs}. For example, Rashid et al.~\cite{rashid2012cost} observed significant overhead when switching between displays and classified potential factors~\cite{rashid2012factors}, such as content coordination and input directness. This overhead is also apparent when switching between content displayed on an AR-HMD and a physical screen~\cite{hubenschmid2023smartphone}. We take inspiration from these works and insights, to design a system that leverages the strengths of different interface types, while considering such potential drawbacks.

\section{\system: \systemLong}
    \system is a hybrid authoring tool for creating complete AR-based human-robot collaboration interfaces, integrating on predefined robot behaviors and task-related information (e.g., assembly instructions, bill of materials). 
    We exemplify \system based on a real scenario informed by our main industry partner (\lego), 
    which involves the assembly and disassembly of plastic injection molds. An injection mold can weigh up to 1.000\,kg and consists of a great number of different parts (e.g., metal plates, bolts, pins, o-rings) that must be put together in the correct sequence using a variety of different tools. 
    To lower physical strain and risk of injury, such assembly processes can be supported by robot arms, where the robot might be responsible for repetitive steps, such as fastening bolts or the robot can act as a flexible fixture for the partially-assembled mold to support ergonomic posture of the operator, who then inserts pins, attaches o-rings, or applies grease.
    Such collaboration requires fluent communication, e.g., the robot notifying the operator, when it is waiting for components to be attached, as well as joint workspace awareness between operator and robot, e.g., to prevent collisions. This can be supported through situated information visualizations with AR (e.g., \cite{lee2023design, martins2022augmented}).
    To ensure applicability in real-world scenarios, we must accommodate different workspaces, mold designs, and operator preferences. 
    This calls for a flexible authoring solution that allows for ad-hoc modifications of the system configuration, which we propose to support with a hybrid user interface.
    
\begin{figure}[h]
        \centering
        \includegraphics[width=\columnwidth]{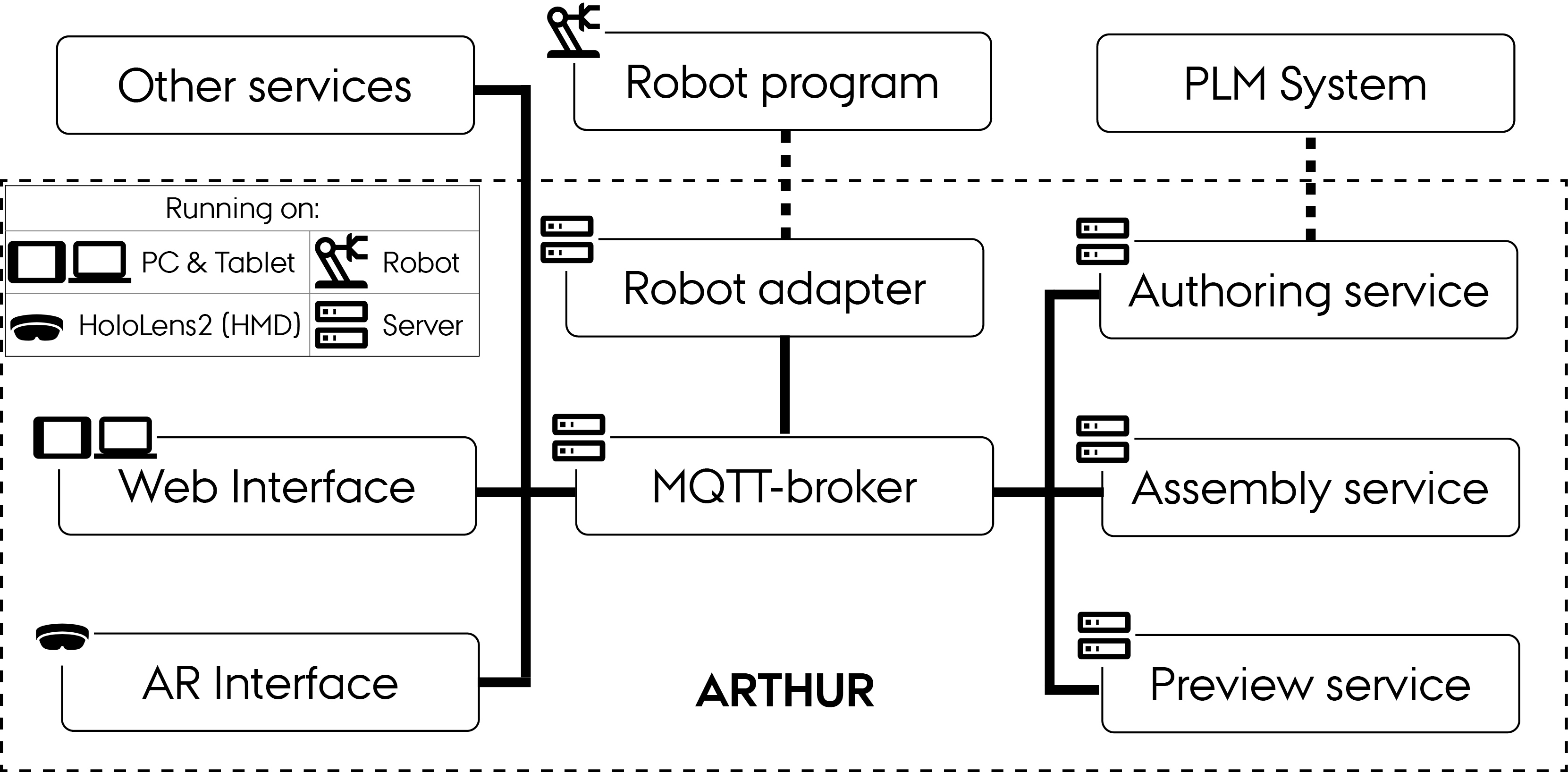}
        \caption{Figure showing all interfaces and services and which device they live on. The boundary indicated by a dotted line highlights what is included in \system and what must be provided, e.g., the robot program is not created in \system or is a direct part of it but communicates with the system through the robot adapter service.}
        \label{fig:interfaces-and-services}
    \end{figure}

    The system setup in our lab is comprised of three main hardware components: an HMD (HoloLens 2), a Robot (UR5e~\cite{ur5e}), and a server (Asus PN51-E1). The user interacts through two main interfaces (Web interface, AR Interface) that were built with Vue and Unity respectively, and rely on a range of services, as illustrated in \autoref{fig:interfaces-and-services} (see project repository\footnotemark[\repoVal] 
    for further technical details).
    The web interface supports efficient authoring of feedback and actions using traditional UI elements (e.g., menu selection, text input) on a PC or tablet. The \textit{AR application} presents situated visualizations and allows testing and refining the system as well as manipulating content directly in the workspace. For this, the HMD supports various input modalities, such as gaze, gesture interaction, and speech. 

    \subsection{Authoring Workflow with a Hybrid User Interface}
        We divide the overall workflow with \system into three phases, between which the user can flexibly transition: (1) \phaseOne, (2) \phaseTwo, and (3) \phaseThree. These utilize the web and AR interfaces to different degrees, as is described below and illustrated in \autoref{fig:teaser}.

        In the \phaseOne phase, the general setup is defined on a desktop computer, where we can make full use of the interoperability of several applications (e.g., Product Lifecycle Management systems; PLM) to import existing data (e.g., parts, tools, assembly sequences), 
        as recommended by prior work~\cite{hubenschmid2022relive}. 
        \begin{figure*}[h]
            \centering
            \includegraphics[width=\linewidth]{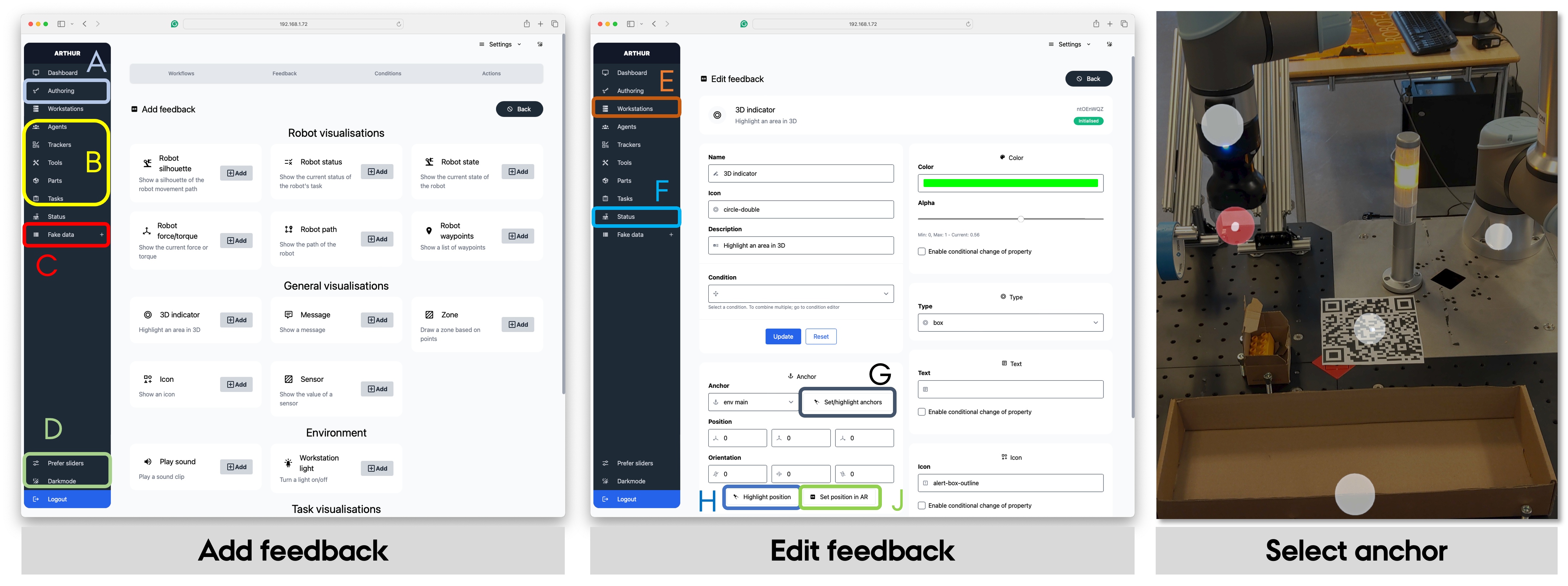}
            \caption{Left and middle show screenshots from the web interface, where the rendering is the same across devices (PC and tablet). Right shows the AR interface after clicking ``Set/highlight anchors'' (G). The pages are similar for all design components (\autoref{fig:building_blocks}), with a page for viewing, creating, and editing feedback, actions, and conditions. A position can also be viewed in AR (H) and set (J). The sub-menus for authoring is found under (A), it is possible to create a new system setup from the ``Workstations tap'' (E), edit agents, trackers, tools, parts, and tasks (B), get a status off all services (F), generate ``fake data'', e.g. path, waypoints, zones, and messages, which can be used to test the appearance of some visualizations in current lack of any real data (C), and finally customize the interface, e.g. by showing sliders for number input instead of raw number input (D).}
            \label{fig:interface}
        \end{figure*}
        Considering the example of the injection mold case, one would start by creating a new configuration (under the Workstation tab, \autoref{fig:interface}, E). Agents can then be added, which in our scenario are a robot (UR5e) and an operator (see \autoref{fig:teaser}). 
        Next, a tracker (QR-code) is added to anchor the robot and virtual content (\autoref{fig:interface}, Right).
        Then information about the procedure (e.g., task sequence, tools, and parts) can be added. In our case, these details were exported from the Siemens TeamCenter PLM system, converted from \textit{.XML} into our \textit{.json}-format\footnote{The \textit{.json}-format conversion is similar as described by \citeauthor{park2023digitalizing}~\cite{park2023digitalizing}; sample scripts for conversion are available in the project repository.} and then imported into the system.
        
        With all the basic information in place, the authoring of AR content can begin (\autoref{fig:interface}, A) by creating and configuring the initial set of feedback, actions, and conditions. 
        The \textit{web interface} offers a great variety of design components as building blocks for AR-based HRC workflows, which are further detailed in the next section. \autoref{fig:interface} (left) shows a list of possible feedback that can be created and \autoref{fig:interface} (middle) shows how properties of a feedback component can be edited. The views for \actions and \conditions are similar. 
        In our injection mold scenario the AR content could be: \textit{robot path} that highlights the expected movement of the robot, \textit{robot state} visible when stopped, \textit{robot status} visible when running showing the current task it is working on and its progress, \textit{step model highlight} for the operator tasks showing a hologram of where parts are to be placed, maybe a \textit{step model highlight} for the robot's tasks with reduced opacity, \textit{tool} and \textit{part highlights} showing what tools and parts to use when relevant, and finally a warning \textit{sound} when the robot initiates a movement.   
        Updating the labels, names, icons, and descriptions for design components in this phase is preferable, as the interface offers the best overview and includes a physical keyboard.

        Once the general configuration is made, the user equips an AR-HMD and transitions to the robot workspace to begin the \phaseTwo phase (see \autoref{fig:teaser}, ``Refinement''), for refining content position and appearance, and testing interactivity. Here the \textit{AR interface} displays the previously authored visualizations in situ and allows editing through hand-based 3D manipulation, such as repositioning objects (e.g., content anchors, see \autoref{fig:interface}, right) within the robot's workspace via mid-air gestures.
        Simultaneously, more complex changes or fine adjustments to the workflow may still be made through the previously described \textit{web-interface}.
        This can now be accessed through a tablet that is mounted in the near periphery of the robot workspace or hand-held, whereby touch input is supported in place of mouse and keyboard.
       We hereby capitalize on the \textit{hybrid user interface} concept, benefiting from immediate in-situ visualization in AR, while effectively authoring changes on the touchscreen. Thus, contrary to using Unity or other tools that require some reloading or rebuilding of the application, changes are immediate visible in the \textit{AR interface}.

        In our injection mold scenario we may want to align the \textit{task image panel} such that it is just in front of the workpiece on the table, as illustrated in \autoref{fig:teaser} (see ``Refinement''). Some elements cannot be re-positioned, such as the \textit{robot path} that is ``attached'' to the robot, and the \textit{step model highlight} as the position for each step in an assembly sequence is defined in the Bill of Process (BoP). However, we can verify that the robot and root element of the assembly sequence are correctly positioned in relation to the tracker (physical QR-code) and adjust the offset, if needed. All parts and tools defined in the Bill of Material (BoM), e.g., a hammer or a brush and grease can, can also be localized, so that they may be highlighted when relevant. \system can be extended with dynamic localization of tools and parts (e.g. through a vision system), though this is not currently not included in the system.
        We may also wish to adapt the color, line widths and sizes of visualized content (e.g., robot path) for good visibility and minimal occlusion. Finally, we test that the physical buttons in our setup, which can be used to start/stop the robot and mark tasks in the sequence as completed, work as expected.

        Lastly, in the \phaseThree phase, the user tests the fully authored workflow on the HMD by performing the task in collaboration with the robot. The authored system may also be handed over to a different user, who may wish to make ad-hoc changes (e.g., to address a spontaneous change in procedure or accommodate personal preferences) before completing the job.
        
        Our hybrid user interface approach allows the user to fluidly switch between devices, opting for the most convenient technology in each step of the authoring, testing, and deployment procedure, thereby speeding up the design-deploy-refine process. 
        To give a small but very frequently ocurring example: naming design components in \phaseOne phase requires text input, which is most effectively supported with a physical keyboard on the PC. Typing is required less often in \phaseTwo phase, when components need to be renamed or new ones added. This is then supported through touch typing on the tablet, which -- though inferior to typing on a physical keyboard -- is preferable to typing in mid-air using the HMD~\cite{lunding2024robovisar}. In contrast, while content positions can be efficiently initialized on the PC, its refinement is left for the \phaseTwo phase, where the HMD supports in-situ inspection of content alignment with the physical workspace. Here, coarse re-positioning of content is supported through direct manipulation, while the tablet interface offers fine-tuning controls.

        \subsection{Design Components}
\begin{figure*}[ht]
                \centering
                \includegraphics[width=\linewidth]{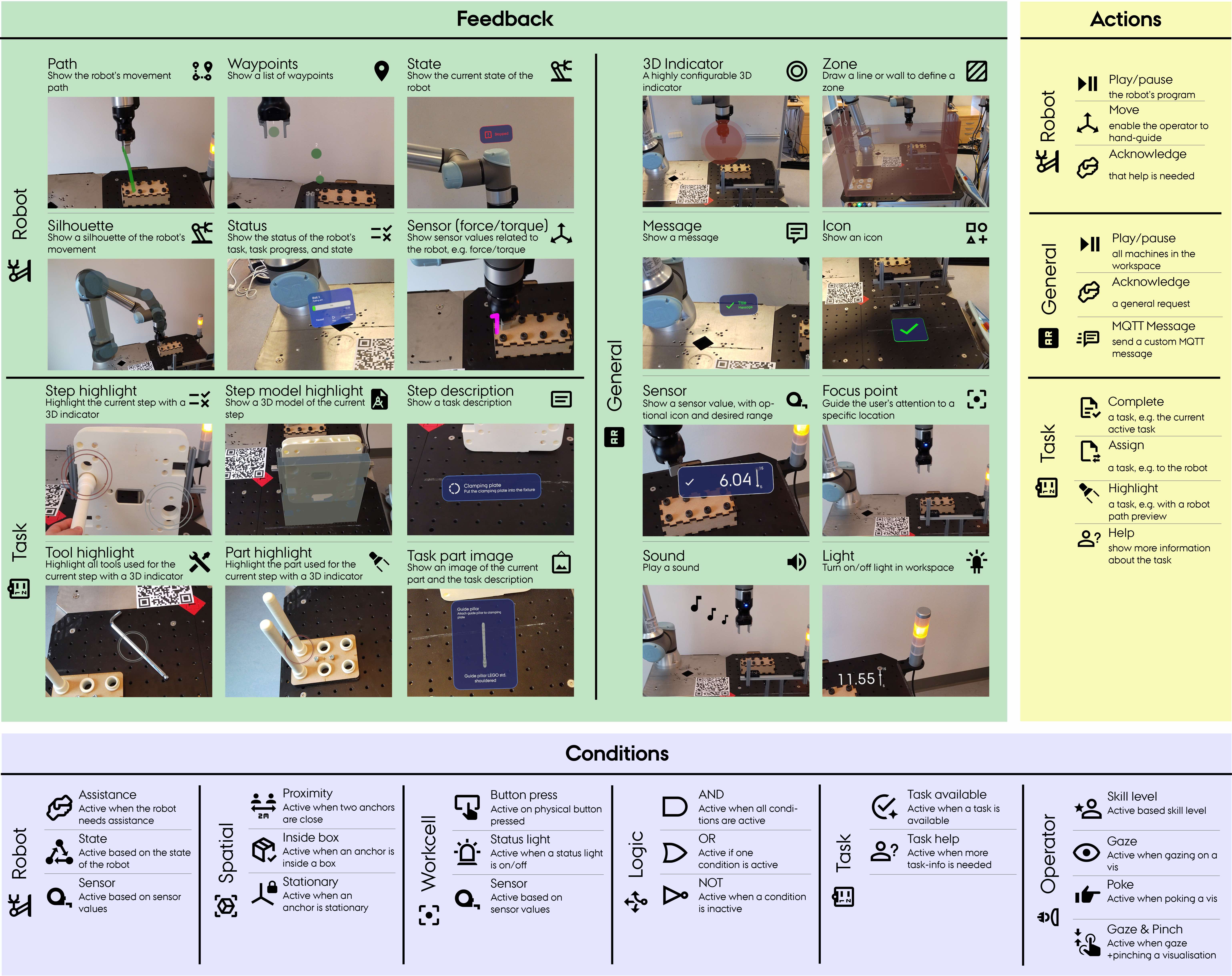}
                \caption{Overview of all \feedback, \actions, and \conditions currently implemented in \system.}
                \label{fig:building_blocks}
            \end{figure*}

            \system{} offers three kinds of design components as building blocks (see \autoref{fig:building_blocks}): \feedback communicates the state of the robot and system to the user (e.g., as in-situ visualizations); \actions{} allow the user to control the robot and system; and \conditions serve to automatically trigger \actions or dynamically adjust \feedback properties. All components can be individually customized through properties (e.g., width and color of robot path visualization), which are instantly synchronized across all connected interfaces. In addition, \system{} supports \textit{trackers and anchors} to easily and consistently place content in the real world, and \textit{task-related data} (see Section \ref{sec:otherComp}). 

            \subsubsection{\Feedback}
            \label{sec:feedback}
               
                \system{} currently supports \nVisualisations kinds of \feedback (see \autoref{fig:building_blocks}) to inform users about the state of the robot and system. While we currently focus on visual \feedback (e.g., visualizations), we aim to explore other modalities afforded by AR HMDs (e.g., audio, haptics) in future work. We group \feedback into visualizations that concern the \textit{robot}, the \textit{task}, or \textit{general} purpose.

                \textit{Robot}-related \feedback includes visualizations about the robot's movement path, waypoints, silhouette, state, or sensor readings. In addition, users can add a visualization about the current task status. 
                While most of the data can be gathered directly from the robot in real time, some feedback (e.g., path, silhouette) requires estimations about future movements. 
                We support this through the \textit{Preview service}, which is described in Section \ref{sec:services}.
                
                \textit{Task}-related feedback guides the user through the workflow. We support step-by-step instructions (e.g., task part image) and highlights (for task, task model, tool, or part), allowing users to quickly identify relevant components. Further, the overall task status allows them to monitor progress and coordinate activities with the robot.

                Lastly, \textit{general} \feedback enables to customization of the user's workspace and environment. This includes the capability to show 3D indicators, icons, or zones, thus enabling support for a wide range of scenarios (see Section \ref{sec:scenarios}). For example, a 3D indicator can be turned into a virtual button by utilizing its ``poke'' condition or to create a ``stay-out area'' where the user is not supposed to be inside while the robot is moving. Likewise, icons can be used to convey the state of the system (e.g., if a sensor value is at its desired level or an object is placed correctly). Finally, physical lights within the workspace and spatial audio can be used to provide feedback.  
                            
            \subsubsection{\Actions}
            \label{sec:actions}

                \Actions are ways for the user to communicate or express commands to the system. Similar to prior work (e.g., \cite{suzuki2022augmented}), we support input modalities commonly available in current AR HMDs, such as gaze, speech, and gestures, as well as relative object position (e.g., proximity). 
                While \actions vary greatly by usage scenario, we implemented a basic set of \nActions~common \actions based on our related work analysis (see \autoref{fig:building_blocks}). We again differentiate actions that concern the \textit{robot}, \textit{task}, or are \textit{general} purpose.

                \textit{Robot} \actions allow the user to play or pause the robot program, send a confirmation, or trigger move mode (i.e., hand-guiding). It should be noted that these \actions might not be applicable or possible on all types of robots and in every context. \textit{Task}-related \actions give the user the opportunity to control the sequence of tasks, for example by manually confirming the completion of a task, reassigning it to a different agent (i.e., task scheduling), or selecting a specific task to view more information about it. Finally, \textit{general} \actions work on a broader scale and currently include a general acknowledgment \action, a global play/pause function (e.g. for all machines in the workspace), and transmit custom MQTT \cite{mqtt} messages, thus allowing the operator to send input to other connected systems. 

            \subsubsection{Conditions}
            \label{sec:conditions}
                \Conditions can be used to control when, where, and how \feedback appears or to trigger \actions. We have implemented \nConditions \conditions in 5 different categories: \textit{spatial}, \textit{operator}, \textit{robot}, \textit{environment}, \textit{task}, and \textit{logic} (see examples in \autoref{fig:building_blocks}).
                To illustrate a potential usage scenario: We might apply \textit{Spatial} \conditions to hide detailed visual information for a workspace the operator is not currently attending to by detecting when they are more than three meters away (\textit{proximity condition}). Another example is the \textit{robot assistance} \condition which can be triggered to alert the operator to this workspace when needed, e.g., because the material dispenser is empty.

                \Conditions are generally created through the \textit{web interface}. However, some \feedback types and \actions imply \conditions, which are automatically created.  For example, the creation of interactive visualizations, such as the \textit{3D Indicator} and \textit{Task model highlight}, automatically includes a corresponding gaze, gaze+pinch, or poke \condition.
                All \conditions can further be customized: they are either active or inactive and some provide additional parameters. For example, gaze, gaze+pinch, and poke refer to the respective visual element.

            \subsubsection{Trackers and Anchors}
                To register virtual content in the real world (e.g. the robot movement path) it is necessary to align the coordinate systems of the robot and HMD. Furthermore, it is desirable for some virtual content to be fixedly situated in the environment. This can be authored using \textit{trackers} and \textit{anchors}.
        
                \textit{Trackers} are used to localize the HMD in the world. Currently, this is supported through QR-codes that can be applied to physical surfaces and scanned with the HMD (HoloLens2). 
                All trackers have an \textit{anchor} attached, which then can be used to fix virtual content and position robot(s) in relation to it.
                Additional \textit{anchors} can be created to specify fixed content locations, e.g., at the base, joints, and tool-center point of the robot. For those familiar with Unity3D or similar concepts: a tracker can be seen as a game object with an anchor at its local origin (0,0,0). Additional anchors can be created in relation to these tracker anchors.

                By default, anchors are also attached to the user's hands and head, which is tracked by the HMD. This can serve e.g., to define a \textit{proximity condition} that activates based on a distance-threshold between two anchors (e.g., distance between user's hand and end effector of the robot).
                
            \subsubsection{Other Components}
            \label{sec:otherComp}
                \system allows the definition of agents (i.e., operators and robots)
                with specific attributes. For example, operators have a skill level that may modulate the level of detail of visualized information. Robots have a type, (e.g., UR5e, UR10e, KUKA iiwa), tools (e.g., gripper),  and a position relative to a tracker in the environment. This serves to instantiate and situate the corresponding model of the robot. 
                
                Further assembly-relevant data includes the tools and components (BoM), and tasks (BoP). These are imported from external sources, e.g., a PLM system, such as SIEMENS Teamcenter.(A similar approach was pursued by S{\"a}{\"a}ski et al. \cite{saeaeski2008integration}.)
                To facilitate correctly situated visualizations based on this data, the location of tools, parts, and tasks can be specified as part of the authoring process, e.g., by relating it to anchors or by adding a custom Tool Recognition Service (e.g., using vision-based scene segmentation approaches).
                The supported assembly procedures were informed by prior work \cite{park2023digitalizing,saeaeski2008integration} and our industry partners in MADE FAST. 
                
        \subsection{Services \& Extensibility}
        \label{sec:services}
            \system offers multiple services that support the two interfaces and various aspects of HRC. An overview of the system architecture is given in \autoref{fig:interfaces-and-services}.

            The \textit{authoring service} represents the system's core and is responsible for handling and storing all changes to the configuration of the authored setup. Thus the authoring service is required for \system to function. 
            
            The \textit{assembly service} is an optional service responsible for bookkeeping during an assembly process, where it loads the assembly sequence from a database shared with the authoring service and manages the task scheduling (task assignments and completion status). 
            
            The \textit{preview service} is an optional service that implements a simple way of getting robot motion previews, used for path and silhouette feedback, without the need for a simulation. The service automatically records the tool-center points (TCP) and joint angles of the robot when it is doing a task. This data can then be replayed on request, e.g., for displaying the robot movement path to convey motion intent and planned actions. 
            
            A \textit{robot adapter service} is responsible for 
            handling the communication with the robot, like retrieving joint angles and TCP, and informing the robot about tasks that it must perform. Currently, \system supports a Universal Robots (UR)~\cite{ur5e} adapter. The UR-adapter uses the RTDE-interface \cite{rtde} to get data from the robot and an XML-RPC server from which the robot program can retrieve information about its next task and send progress updates. The robot program is not a part of \system as it will be unique to every context. Communication between the robot and its program is intended to happen through the robot adapter.
            Additional robots can be added in the future. From the perspective of \system, this is done by implementing a new \textit{robot adapter service} for that specific type of robot. The adapter must implement the minimum required functionality, e.g. data from the robot like joint angles, TCP, and robot state and data to the robot, such as which task(s) it may perform. The details, likely to change over time, can be found in the repository\footnotemark[\repoVal]. The robot adapter can be ROS-based if applicable.
    
            To handle communication between all user interfaces and services, \system relies on MQTT~\cite{mqtt}. This allows clients to exchanging JSON-encoded messages via publish and subscribe to topics via an MQTT-broker~\cite{mosquitto}. The integration of additional services and systems is well-supported by MQTT.
            For example, the system can be extended by an additional 'safety-zone' service, as described in Section \ref{sec:scenario_1} or to communicate with ROS-services by using a ROS-MQTT-bridge.
            
            If additional design components (i.e., \feedback, \actions, and \conditions) are needed, \system allow developers to extend the system with those.
            This is a fairly simple process, which requires changes to two parts of \system:
            (1) Basic information about the component: name, icon, description, and a list of all properties 
            must be added to the \textit{authoring service}. The following 12 properties are currently supported: \textit{boolean, integer, float, string, anchor, pose, vector3, condition, color, agent, enum, multi select enum}.
            As an example, the path visualization has the following properties: agent, width (float), and color.
            (2) The \textit{AR interface} 
            must also be updated with the actual implementation. Each type of component has a specific interface that must be implemented, which generally comprises three methods: init, cleanup, and ListOfProperties. By implementing this interface, the new design component gains access to all data in the system. 
            More details are available in the documentation\footnotemark[\repoVal].
        
\section{Evaluation of \system}
We investigate the potential of \system through \textit{demonstration} (Section \ref{sec:scenarios}) and verify the suitability of our design in a qualitative \textit{usage evaluation} with experts (Section \ref{sec:experteval})~\cite{ledo2018evaluation}. \rev{Extensive related work have found AR and AR-HRC to be useful~\cite{suzuki2022augmented,lunding2023ar,rosen2019communicating,lunding2024robovisar,hietanen2020ar,andersen2016projecting}, thus this evaluation focusing on the capabilities and usage evaluation of \system.}

\subsection{Demonstration of Application Scenarios}
\label{sec:scenarios}
    To exemplify the capabilities of \system, we explore three application scenarios that involve: \textit{replicating a system} \cite{hietanen2020ar}, creating an environment for \textit{comparing visualizations}~\cite{arevaloarboleda2021assisting,cleaver2021dynamic,cogurcu2023augmented,gruenefeld2020mind}, and \textit{displaying sensor values}~\cite{defranco2019intuitive,fuste2020kinetic,renner2018wysiwicd}, as described below and illustrated in our supplementary videos\footnotemark[\videoVal].
    
    Scenarios were chosen based on replicability (i.e., all central components were described in sufficient detail), and as they represent major current directions in HRC research, or address central pain points in the manufacturing industry.
    The chosen scenarios highlight the challenges of current authoring approaches (e.g., \cite{lunding2024robovisar}), as they require a mixture of in-situ (e.g., placing objects in 3D, choosing color and opacity of virtual objects, defining suitable thresholds and distances) and ex-situ authoring (e.g., importing data). \system was not specifically designed for any of these scenarios and may be more widely applied.
    
    \subsubsection{Scenario 1: Replicating a System}
    \label{sec:scenario_1}
\begin{figure}[h]
            \centering
            \includegraphics[width=\columnwidth]{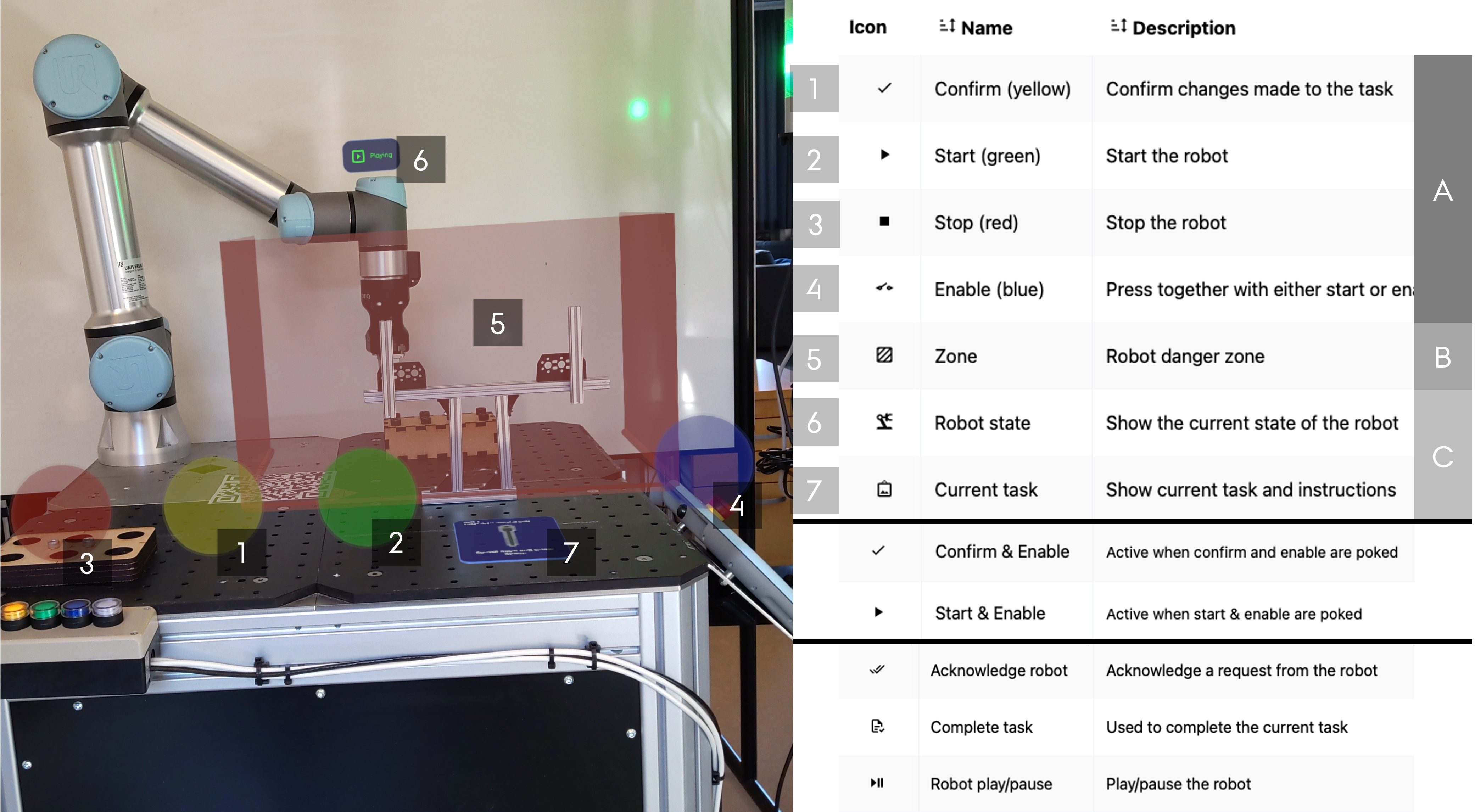}
            \caption{Scenario 1 \cite{hietanen2020ar} replicated in \system. The left part shows the view from the AR interface, with each design component annotated with a number. The right part depicts a list of feedback from the web interface, with corresponding numbers annotated, a list of created conditions, and finally a list of actions.}
            \label{fig:scenario_1}
        \end{figure}

        In this scenario, we replicate the HoloLens setup as described by Hietanen et al.~\cite{hietanen2020ar}. The original setup consist of six~visual elements, though seven are used in \system to achieve the same result (see \autoref{fig:scenario_1}): 
        (A)~touching four differently colored spherical objects in the workspace allows the user to \textit{start} (green) and \textit{stop} (red) the robot, \textit{confirm} (yellow) changes that have been made to the task, and \textit{enable} (blue) the start and confirm buttons, (B)~a dynamically-updated safety-zone visualized as a semi-transparent wall around the robot, finally (C)~a graphical box with an image and text showing the robot's current status and instructions.

        The setup can be replicated in \system as follows:
        (A)~four \textit{3D indicator} visualizations with appropriate shape (sphere), color, and positioning. Each \textit{3D indicator} has an associated \textit{poke \condition}, which allows for interactivity through mid-air gestures. The \textit{stop button} (red) was set to activate a \textit{robot play/pause \action}. The \textit{start button} (green) informs the robot that it may initiate its next task, and is thus set to trigger a \textit{robot acknowledge \action} when poked simultaneously with the enable (blue) button. This is achieved using an \textit{AND} condition.
        (B)~a \textit{zone visualization} can be used to display a semi-transparent wall around the robot. While \system currently does not explicitly support automatically update of safety zones, it is easy to extend the system with a service that calculates and continuously publishes a series of points around the robot along with a zone-id through MQTT.
        Finally, the \textit{confirm button} (yellow) triggers a \textit{complete task \action} through a simultaneous \textit{poke} of this and the enable (blue) button.
        Lastly, (C)~is split into two elements: a \textit{task image visualization} which shows a panel with a description of the next task, retrieved from the task description, and a \textit{robot state visualization} that shows the current state of the robot (i.e., playing, paused, stopped).

    \subsubsection{Scenario 2: Comparing Visualizations of Prior Work}
\begin{figure}[h]
            \centering
            \includegraphics[width=\columnwidth]{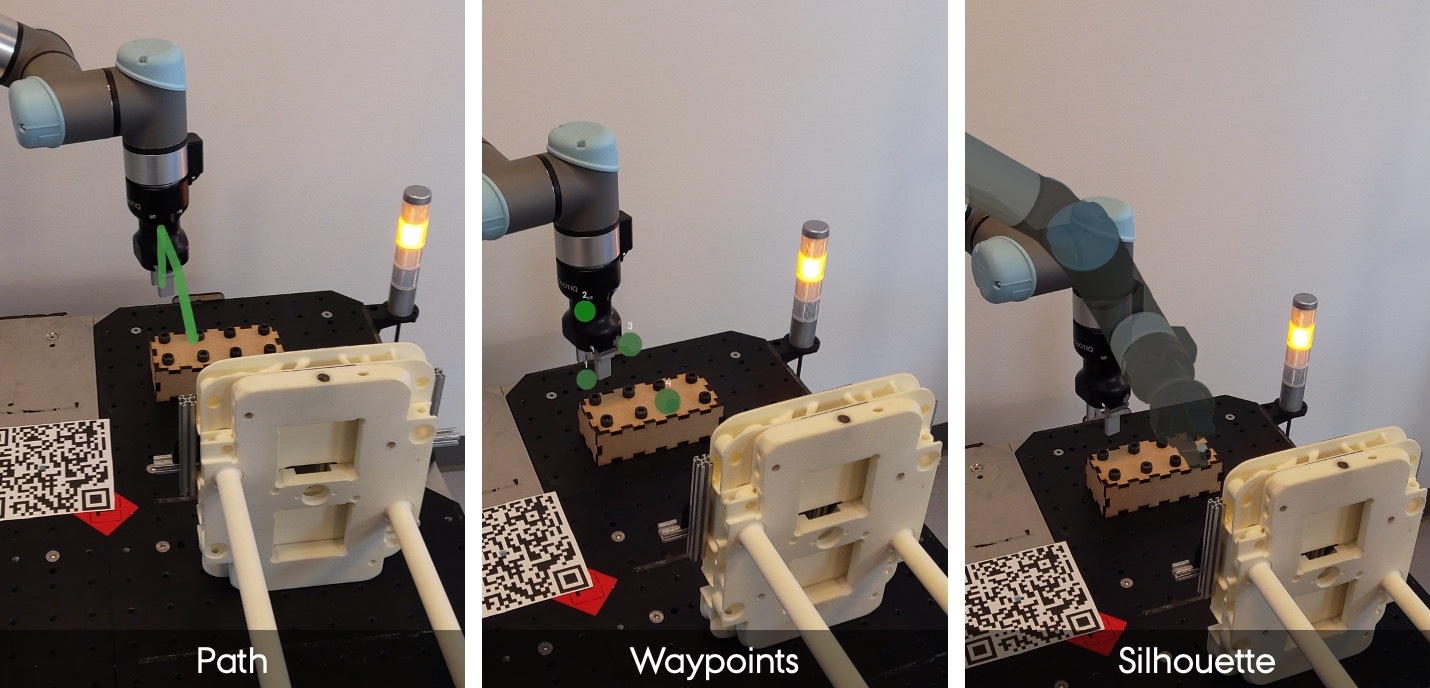}
            \caption{Setup for comparing previews of robot motion. Similar setups can be found in \cite{cleaver2021dynamic,cogurcu2023augmented,rosen2019communicating,arevaloarboleda2021assisting}.}
            \label{fig:scenario_2}
        \end{figure}

        There is a considerable amount of related work exploring the benefits of individual visualization types, e.g.~\cite{arevaloarboleda2021assisting, cleaver2021dynamic,cogurcu2023augmented,gruenefeld2020mind}. However, the distinct physical setups and application scenarios make findings difficult to compare.
        As \system already contains an extensive collection of \feedback, \actions, and \conditions, it provides a good starting point for recreating and comparing variants of visualizations from prior research.
        For example, we can recreate various visualizations for robot movement intent, as illustrated in \autoref{fig:scenario_2} for direct comparison in \system.
        Our hybrid authoring approach allows users to easily realize such visualizations following our three phases: In \phaseOne phase, users can set up a general workspace with different visualization and interaction elements, to switch between these visualizations on the fly (similar to Scenario 1). In \phaseTwo phase, they can fine-tune these visualizations on a tablet and observe the output in situ with the AR-HMD (e.g., making sure the color of visualizations are appropriate). Finally, in the \phaseThree phase, users can seamlessly test, switch, and compare the visualizations.

    \subsubsection{Scenario 3: Sensor Values}
        Integrating and visualizing sensor values (e.g., from robot or attached tools) is a common scenario for assisting the user during HRC, e.g. ~\cite{defranco2019intuitive,fuste2020kinetic,renner2018wysiwicd}. This might be most relevant when programming the robot, but can also be useful during operation to verify that the robot operates within the desired limits. For example, prior work has explored battery indicators~\cite{renner2018wysiwicd} or displaying a scale to inform about the amount of loaded material~\cite{fuste2020kinetic}.

\begin{figure}[h]
            \centering
            \includegraphics[width=.8\columnwidth]{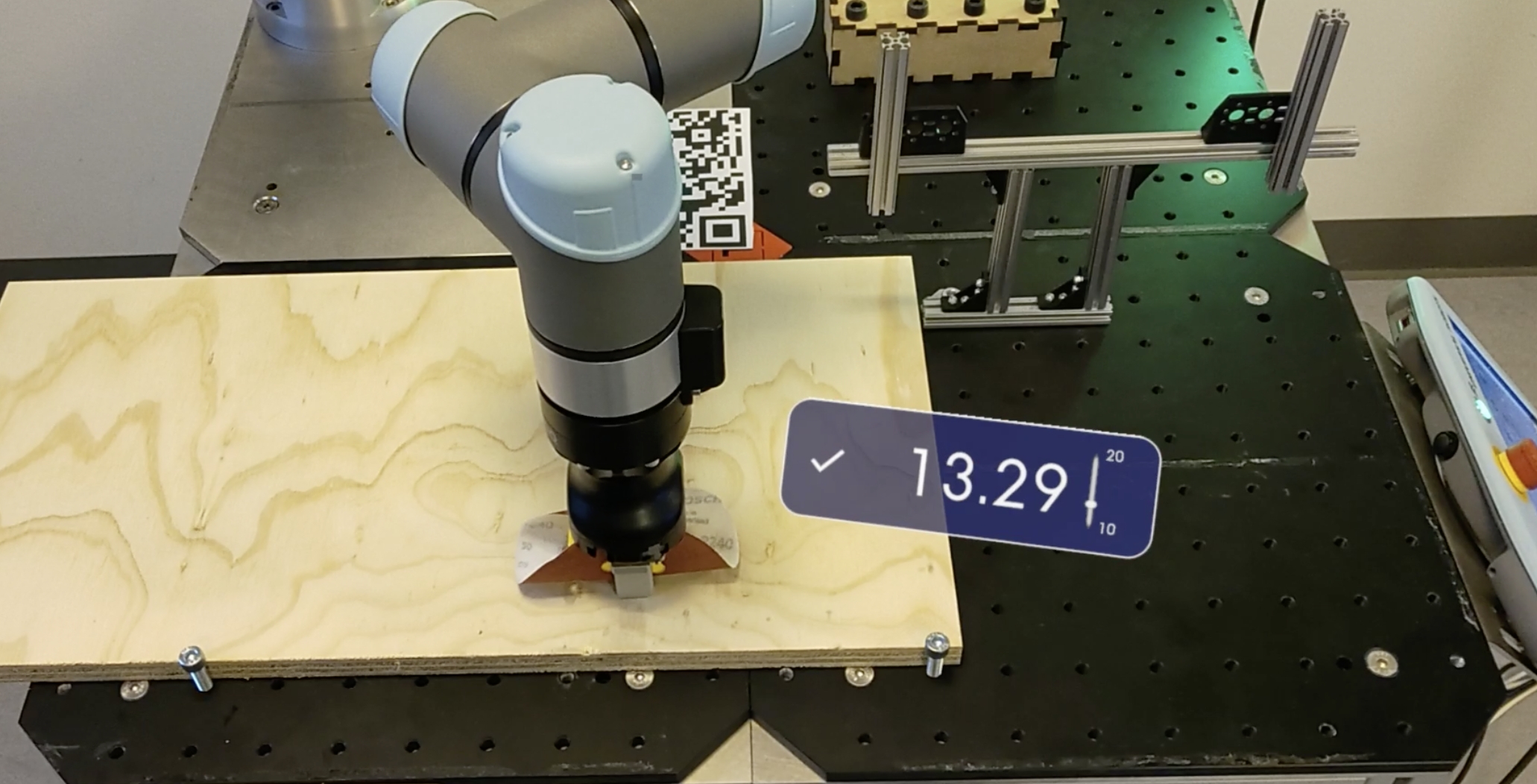}
            \caption{Pressure-sensing sanding setup \cite{defranco2019intuitive} replicated in \system.}
            \label{fig:scenario_3}
        \end{figure}

        \system{} can be used to easily replicate similar setups, such as the work by De Franco et al.~\cite{defranco2019intuitive}, where the amount of pressure applied during sanding is visualized in an AR interface (see \autoref{fig:scenario_3}). Using our set of design components, we can create a \textit{sensor visualization} and connect it a digital scale visualization via our MQTT-interface.
        The general setup and configuration (e.g., reading documentation on the scale, setting up MQTT topics) is best performed on a desktop during \phaseOne. 
        In contrast, the position of the \textit{sensor visualization} can be either attached to an anchor within the scene or placed through mid-air gestures and fine-tuned in-situ -- thus, this step is best done using an AR-HMD and tablet during \phaseTwo. Lastly, users can immediately test and validate the visualization by switching to the \phaseThree phase.
        
\subsection{Usage Evaluation with Experts}
\label{sec:experteval}
    To verify the suitability of the supported authoring functionality across interfaces, we performed a usage evaluation~\cite{ledo2018evaluation} with \rev{five} participants with various expertise within AR and AR-HRC.
    Our goals were to (1)~observe and assess how \system is used to author a holistic HRC process, (2)~examine the qualitative usability and utility of \system, and (3)~validate our hybrid approach of distributing the authoring across multiple phases and devices.
        
    In particular, we aimed to investigate whether we had leveraged the potential of AR for the given task, hence our recruited participants had relevant expertise in AR development and robot interaction.
    All participants were male, between 22 and \rev{36} years of age, and identified as researcher (P1), PhD-student (P2), graduate student (P3), and software developer (P4, P5).
    On average, participants rated their experience with head-mounted AR as high and working/interacting with robots as medium but indicated little to no experience with hybrid user interfaces and robot programming.

    \subsubsection{Procedure}
        After welcoming participants to the lab, participants gave their written consent for voluntary participation and data collection, and filled a demographic questionnaire. We then introduced the study task, which was to replicate the system described in Scenario~1, and explained the different phases of \system{}. We explained the hybrid authoring approach, which included a walkthrough of all sub-menus in the web-interface (see \autoref{fig:interface}), an exemplary creation of a message (\feedback), workstation button (\condition), and play/pause robot (\action) on a desktop computer, followed by inspecting the output on the HMD and starting a dummy robot-program by triggering the play/pause robot \action. We also demonstrated the cross-device interaction by starting the 3D placement of the created message (\feedback) on the tablet, then switching to the AR-HMD to place the message within the workspace, and switching back to the tablet to confirm its position. Once participants felt comfortable with the system, we explained Scenario~1 by showing them the intended outcome (see \autoref{fig:scenario_1}) and describing all visual elements and their expected functionality.
        
        Participants then started the authoring process at a desktop computer (\phaseOne phase), before transitioning to the physical setup with the AR-HMD and Tablet (\phaseTwo phase), and then finally testing their setup (\phaseThree phase). Once they completed the task, they were provided with additional tasks to further explore the system, e.g., creating 3D step instructions, adding auditory feedback to button interactions, and only showing the zone when the robot was active. They were then invited to try out their additions and freely explore the system further before the session concluded with a semi-structured interview.
        Each session lasted 75 minutes, whereby the explanation of scenario one and the three phases took up 25-30 minutes.

    \subsubsection{Main Usage Evaluation Findings}
        \rev{We evaluate \system through thematic analysis of the experimenter's written protocol of each study session that was further refined by inspecting video and audio recordings, including transcripts from interviews.} In summary, \rev{all participants where able to complete the study task by replicating the setup presented in scenario 1, including adding and configuring design components of their choice.} Furthermore, all reported that \system works well and is easy to use, that the options were plentiful, and that distributing tasks across different devices in the authoring process is beneficial:
        \rev{\textit{``The combination of a decent input/menu device (tablet or PC) and immediate feedback on the Hololens is great.''} - P5}

        \rev{The participants found it great to have a tablet as a menu, instead of virtual menus on the HoloLens. 
        It was also highlighted as a strong positive by P4 and P5 that AR assembly instructions are not authored per-instruction, but by loading the BoP and use generalized visualizations, as this have the potential to drastically reduce the time needed for authoring of step instructions, especially with slightly different tasks.}
        
        Participants appreciated the hybrid authoring approach, for example allowing them to use AR for coarse object positioning and the tablet for fine-tuning: \textit{``It was easy to use the tablet simultaneously with the HMD - they complement each other well.''} - P3. Four participants highlighted the benefits of a mixture of devices: \textit{``The utility of the PC increases as the system becomes more complex. But for smaller, simpler setups, the iPad is great, as it is much more nimble, and you can be in-situ with the [AR HMD] on.''} - P1. Participants reported no issues when switching between authoring phases, but valued that they could: \textit{``quickly do things here [on the computer] and then go over to place stuff [in-situ]''} - P2.
        \rev{The context switching were not perceived as a problem, as: 
        \textit{``You can transition between the tablet and Hololens so easy. You just press [on the tablet] and then it updates in AR. No need to wait for a configuration to load or similar, you just click and then it updates.''} - P4}

        While the participants overall were pleased with \system, they noted some usability issues regarding the AR authoring workflow. When setting a position in AR, they unintentionally rotated the object. Here, participants suggested a rotation lock or separating the 3D manipulation tasks. Furthermore, all \feedback is currently hidden while positioning an object in AR, which reduces visual clutter, but also makes it difficult to align objects in relation to each other. \rev{A solution could be to reduce the opacity of all visualizations or add a button for toggling visibility. P4 also asked for better ways to align content, which could be achieved with 'snapping', copying the position of other objects, and enable the user to constrain the movement to specific axis.}

\section{Discussion}
       While prior research presents custom implementations of AR guidance or feedback for robot behavior \cite{andronas2021multi, ganesan2018better, hietanen2020ar}, these do not permit modification by the end-user in situ. Authoring AR-based HRC workflows with existing tools on "traditional interfaces" (e.g., designing content in simulated workspaces on desktop PC systems) remains a challenge for various reasons, such as discrepancies between the development environment and the deployed application.
       Recently proposed in-situ authoring tools (e.g., \cite{lunding2024robovisar}) attempt to close this gap, but report limitations of usability due to mid-air interaction modalities. 
       Prior work in the field of visual analytics has indicated that a hybrid approach (i.e., combining heterogeneous devices) allows users to switch between appropriate interfaces as they see fit. However, this was mostly studied in isolation, i.e., either as a discrete switch (cf.~\cite{hubenschmid2022relive}) or simmultaneous use of both interface (cf.~\cite{hubenschmid2021stream, langner2021marvis, reipschlager2021personal}). By combining these use patterns (e.g., discretely switching between the \phaseOne and \phaseTwo phase, but continuously switching between tablet and AR-HMD in the \phaseTwo phase), we showcase such hybrid approaches within a holistic system.
        Initial results from our evaluation indicate that this works well for authoring, as it supports users in transitioning between different tasks.  While there may be increased cognitive load from switching between interfaces (cf.~\cite{rashid2012cost}), we think that the apparent benefits far outweigh possible downsides in our system. However, further studies are obviously necessary to confirm this claim.
        
        \subsection{Limitations of the Expert Evaluation}
            While our initial evaluation demonstrated the general feasibility of our hybrid authoring approach, further studies are needed to investigate the large number of possible design parameters. 
            Also, while we consciously recruited AR experts to explore whether we had leveraged the display technologies and designed the interactive features well, evaluating our system with robotics engineers, UX designers, and workers would contribute another highly valuable perspective. For example, the robotics engineers could judge whether the supported robot operations are representative examples, may identify missed opportunities e.g., regarding relevant robot actions and information to communicate, could explore extending \system to include further robots, tools, and sensors, and might raise technical limitations or issues. \rev{The UX designers, whom is expected to be responsible for authoring AR-HRC systems in the wild, are partially represented in P4 and P5. Getting workers feedback on their ability to make small customization to fit the setup to their personal preferences is also needed.}
            In addition, in-the-wild studies with robot operators and assembly workers are necessary to provide further insights into the real-world applicability of \system and HRC in general~\cite{suzuki2022augmented}.

        \subsection{Beyond Robot Arm Assembly}
    
            While \system focuses on the authoring experience for collaborative assembly with robot arms, we see potential in applying the same system to other areas. For example, support for mobile robots could be added, provided that suitable tracking is supported. Although fiducial markers (e.g., QR codes) already offer decent stability and precision for this scenario, current AR hardware still has significant limitations (e.g., low frequency of marker detection) that make a real-world evaluation difficult on mobile robots. However, if tracking is solved in an environment with mobile robots, a broad range of robot-related \feedback, \actions, and \conditions can be used without any modifications.
            
            One could also imagine using \system for other tasks than assembly, this could e.g. be to supervise robot programming. While it is possible to support robot programming with \system, it will likely require new types of visualizations and data integration to maximize its value.
            Despite our primary focus on robot-supported assembly tasks, we argue that our concepts also apply to a wider range of HRC applications (e.g., robot-assisted inspection, domestic service robots). Our system can be used as a testbed for other researchers, as it accelerates the creation and exploration of AR-supported HRC setups.
            
       \subsection{Configuration Recommendations and Workflows}
            To further improve usability, \system could be extended with a recommendation system that enables the user to add \textit{groups} of \feedback and \actions that play well together, instead of defining everything from scratch. Such recommendations could be generated for individual tasks, or workflows, or more general groups of components that are typically used in existing assembly procedures. Alternatively, these could be based on a shared library of user-defined procedures.
            However, defining, collecting and presenting such recommendations for an HRC setup will require more research to elucidate the needs, suitable strategies, and effective techniques. 
            
            In addition to recommendations, it could be useful to automatically detect all connected sub-systems. For instance, detecting which sensors, buttons, and lights are available in the workstation and only presenting these as options, could help avoid errors by guiding the user to only author \feedback and \actions that are supported by the physical setup. 

\section{Conclusion}
    We propose \system for in-situ authoring of augmented reality guidance for human-robot collaboration through a hybrid user interface. We thereby leverage the familiarity and advanced capabilities of a desktop computer or tablet for menu interaction and an AR headset for 3D visualization and spatial configuration of virtual content through mid-air interaction.
    Hence, our system addresses limitations from previous systems by (1)~extending the possible AR guidance to include assembly instructions; (2)~establishing bidirectional communication between robot and operator (by defining \actions for the operator); and (3)~combining strengths of multiple types of devices and user interfaces to provide a better user experience. 
    
    With this paper, we aim to highlight some of the most central opportunities and challenges of hybrid user interfaces for authoring AR-supported HRC systems. We have implemented the system and shown its capability for replicating a broad range of systems from related work. Lastly, we have shown a path for further exploration of AR-supported HRC by utilizing \system, thereby facilitating the integration of robots to relieve strenuous human manual labor in manufacturing processes.

\bmhead{Acknowledgements}

Thanks go to Roberta, for always doing what it was told to do, no matter how stupid. 
This work was supported by the Innovation Fund Denmark (IFD grant no. 9090-00002B) for the MADE FAST project - part of the Manufacturing Academy of Denmark (MADE) and the \emph{Deutsche Forschungsgemeinschaft} (DFG, German Research Foundation) -- Project-ID~251654672 -- TRR~161.

\section*{Statements and Declarations}

\subsection{Funding}
This work was supported by the Innovation Fund Denmark (IFD grant no. 9090-00002B) for the MADE FAST project - part of the Manufacturing Academy of Denmark (MADE) and the \emph{Deutsche Forschungsgemeinschaft} (DFG, German Research Foundation) -- Project-ID~251654672 -- TRR~161.

\subsection{Conflict of interest}
The authors have no competing financial or non-financial interests to disclose.

\subsection{Ethics approval}
The study was conducted in accordance with the standards and policies for responsible research conduct set by Aarhus University 
and follow national guidelines for ethical research, as well as GDPR rules regarding personal data protection.

\subsection{Consent to participate}
All participants were informed about the study's purpose and procedure and consented to volunteer without financial compensation.

\subsection{Consent for publication}
All participants gave consent for the publication of results based on anonymized data recorded during their study sessions.


\subsection{Materials availability}
Supplementary video clips, e.g., of the demonstrated application scenarios, can be found in the ARTHUR video playlist on Youtube: \video.

\subsection{Code availability}
The source code is available in the ARTHUR project repository: \repo.

\subsection{Author contribution}
All authors were involved in defining the research objective and main research questions. The concept of ARTHUR was jointly developed by R.L., S.H., and T.F.. The system was implemented by the first author (R.L.), who also planned and realized the demonstration of application scenarios. The expert evaluation was planned by R.L. and T.F., while R.L. alone was responsible for preparing and conducting the studies, as well as collecting and analyzing the data. R.L., S.H., and T.F. wrote the main manuscript text, R.L. prepared all figures, and all authors reviewed and revised the manuscript.

\bibliography{references}


\begin{thebibliography}{65}
\ifx \bisbn   \undefined \def \bisbn  #1{ISBN #1}\fi
\ifx \binits  \undefined \def \binits#1{#1}\fi
\ifx \bauthor  \undefined \def \bauthor#1{#1}\fi
\ifx \batitle  \undefined \def \batitle#1{#1}\fi
\ifx \bjtitle  \undefined \def \bjtitle#1{#1}\fi
\ifx \bvolume  \undefined \def \bvolume#1{\textbf{#1}}\fi
\ifx \byear  \undefined \def \byear#1{#1}\fi
\ifx \bissue  \undefined \def \bissue#1{#1}\fi
\ifx \bfpage  \undefined \def \bfpage#1{#1}\fi
\ifx \blpage  \undefined \def \blpage #1{#1}\fi
\ifx \burl  \undefined \def \burl#1{\textsf{#1}}\fi
\ifx \doiurl  \undefined \def \doiurl#1{\url{https://doi.org/#1}}\fi
\ifx \betal  \undefined \def \betal{\textit{et al.}}\fi
\ifx \binstitute  \undefined \def \binstitute#1{#1}\fi
\ifx \binstitutionaled  \undefined \def \binstitutionaled#1{#1}\fi
\ifx \bctitle  \undefined \def \bctitle#1{#1}\fi
\ifx \beditor  \undefined \def \beditor#1{#1}\fi
\ifx \bpublisher  \undefined \def \bpublisher#1{#1}\fi
\ifx \bbtitle  \undefined \def \bbtitle#1{#1}\fi
\ifx \bedition  \undefined \def \bedition#1{#1}\fi
\ifx \bseriesno  \undefined \def \bseriesno#1{#1}\fi
\ifx \blocation  \undefined \def \blocation#1{#1}\fi
\ifx \bsertitle  \undefined \def \bsertitle#1{#1}\fi
\ifx \bsnm \undefined \def \bsnm#1{#1}\fi
\ifx \bsuffix \undefined \def \bsuffix#1{#1}\fi
\ifx \bparticle \undefined \def \bparticle#1{#1}\fi
\ifx \barticle \undefined \def \barticle#1{#1}\fi
\bibcommenthead
\ifx \bconfdate \undefined \def \bconfdate #1{#1}\fi
\ifx \botherref \undefined \def \botherref #1{#1}\fi
\ifx \url \undefined \def \url#1{\textsf{#1}}\fi
\ifx \bchapter \undefined \def \bchapter#1{#1}\fi
\ifx \bbook \undefined \def \bbook#1{#1}\fi
\ifx \bcomment \undefined \def \bcomment#1{#1}\fi
\ifx \oauthor \undefined \def \oauthor#1{#1}\fi
\ifx \citeauthoryear \undefined \def \citeauthoryear#1{#1}\fi
\ifx \endbibitem  \undefined \def \endbibitem {}\fi
\ifx \bconflocation  \undefined \def \bconflocation#1{#1}\fi
\ifx \arxivurl  \undefined \def \arxivurl#1{\textsf{#1}}\fi
\csname PreBibitemsHook\endcsname

\bibitem[\protect\citeauthoryear{Lunding et~al.}{2023}]{lunding2023ar}
\begin{bchapter}
\bauthor{\bsnm{Lunding}, \binits{R.S.}},
\bauthor{\bsnm{Lystbæk}, \binits{M.N.}},
\bauthor{\bsnm{Feuchtner}, \binits{T.}},
\bauthor{\bsnm{Grønbæk}, \binits{K.}}:
\bctitle{Ar-supported human-robot collaboration: Facilitating workspace awareness and parallelized assembly tasks}.
In: \bbtitle{ISMAR'23},
pp. \bfpage{1064}--\blpage{1073}
(\byear{2023}).
\doiurl{10.1109/ISMAR59233.2023.00123}
\end{bchapter}
\endbibitem

\bibitem[\protect\citeauthoryear{Malik and Pandey}{2022}]{malik2022drive}
\begin{bchapter}
\bauthor{\bsnm{Malik}, \binits{A.A.}},
\bauthor{\bsnm{Pandey}, \binits{V.}}:
\bctitle{Drive the Cobots Aright: Guidelines for Industrial Application of Cobots}.
In: \bbtitle{International Design Engineering Technical Conferences and Computers and Information in Engineering Conference},
vol. \bseriesno{86250}.
\bpublisher{American Society of Mechanical Engineers}, \blocation{???}
(\byear{2022}).
\doiurl{10.1115/detc2022-90777} .
\burl{https://doi.org/10.1115/DETC2022-90777}
\end{bchapter}
\endbibitem

\bibitem[\protect\citeauthoryear{Cheon et~al.}{2022}]{cheon2022robots}
\begin{botherref}
\oauthor{\bsnm{Cheon}, \binits{E.}},
\oauthor{\bsnm{Schneiders}, \binits{E.}},
\oauthor{\bsnm{Diekjobst}, \binits{K.}},
\oauthor{\bsnm{Skov}, \binits{M.B.}}:
Robots as a place for socializing: Influences of collaborative robots on social dynamics in- and outside the production cells.
Proc. ACM Hum.-Comput. Interact.
\textbf{6}(CSCW2)
(2022)
\doiurl{10.1145/3555558}
\end{botherref}
\endbibitem

\bibitem[\protect\citeauthoryear{Pascher et~al.}{2023}]{pascher2023how}
\begin{bchapter}
\bauthor{\bsnm{Pascher}, \binits{M.}},
\bauthor{\bsnm{Gruenefeld}, \binits{U.}},
\bauthor{\bsnm{Schneegass}, \binits{S.}},
\bauthor{\bsnm{Gerken}, \binits{J.}}:
\bctitle{How to communicate robot motion intent: A scoping review}.
In: \bbtitle{CHI '23},
p. \bfpage{17}.
\bpublisher{ACM},
\blocation{New York, NY, USA}
(\byear{2023}).
\doiurl{10.1145/3544548.3580857} .
\burl{https://doi.org/10.1145/3544548.3580857}
\end{bchapter}
\endbibitem

\bibitem[\protect\citeauthoryear{Suzuki et~al.}{2022}]{suzuki2022augmented}
\begin{bchapter}
\bauthor{\bsnm{Suzuki}, \binits{R.}},
\bauthor{\bsnm{Karim}, \binits{A.}},
\bauthor{\bsnm{Xia}, \binits{T.}},
\bauthor{\bsnm{Hedayati}, \binits{H.}},
\bauthor{\bsnm{Marquardt}, \binits{N.}}:
\bctitle{Augmented reality and robotics: A survey and taxonomy for ar-enhanced human-robot interaction and robotic interfaces}.
In: \bbtitle{CHI'22}.
\bpublisher{ACM},
\blocation{New York, NY, USA}
(\byear{2022}).
\doiurl{10.1145/3491102.3517719} .
\burl{https://doi.org/10.1145/3491102.3517719}
\end{bchapter}
\endbibitem

\bibitem[\protect\citeauthoryear{Ganesan et~al.}{2018}]{ganesan2018better}
\begin{barticle}
\bauthor{\bsnm{Ganesan}, \binits{R.}},
\bauthor{\bsnm{Rathore}, \binits{Y.}},
\bauthor{\bsnm{Ross}, \binits{H.}},
\bauthor{\bsnm{Ben~Amor}, \binits{H.}}:
\batitle{Better teaming through visual cues: How projecting imagery in a workspace can improve human–robot collaboration}.
\bjtitle{IEEE Robotics and Automation Magazine}
\bvolume{PP},
\bfpage{1}--\blpage{1}
(\byear{2018})
\doiurl{10.1109/MRA.2018.2815655}
\end{barticle}
\endbibitem

\bibitem[\protect\citeauthoryear{Hietanen et~al.}{2020}]{hietanen2020ar}
\begin{barticle}
\bauthor{\bsnm{Hietanen}, \binits{A.}},
\bauthor{\bsnm{Pieters}, \binits{R.}},
\bauthor{\bsnm{Lanz}, \binits{M.}},
\bauthor{\bsnm{Latokartano}, \binits{J.}},
\bauthor{\bsnm{Kämäräinen}, \binits{J.-K.}}:
\batitle{Ar-based interaction for human-robot collaborative manufacturing}.
\bjtitle{Robotics and Computer-Integrated Manufacturing}
\bvolume{63},
\bfpage{101891}
(\byear{2020})
\doiurl{10.1016/j.rcim.2019.101891}
\end{barticle}
\endbibitem

\bibitem[\protect\citeauthoryear{Dimitropoulos et~al.}{2021}]{dimitropoulos2021operator}
\begin{barticle}
\bauthor{\bsnm{Dimitropoulos}, \binits{N.}},
\bauthor{\bsnm{Togias}, \binits{T.}},
\bauthor{\bsnm{Michalos}, \binits{G.}},
\bauthor{\bsnm{Makris}, \binits{S.}}:
\batitle{Operator support in human–robot collaborative environments using ai enhanced wearable devices}.
\bjtitle{Procedia CIRP}
\bvolume{97},
\bfpage{464}--\blpage{469}
(\byear{2021})
\doiurl{10.1016/j.procir.2020.07.006} .
\bcomment{8th CIRP Conference of Assembly Technology and Systems}
\end{barticle}
\endbibitem

\bibitem[\protect\citeauthoryear{Rosen et~al.}{2019}]{rosen2019communicating}
\begin{barticle}
\bauthor{\bsnm{Rosen}, \binits{E.}},
\bauthor{\bsnm{Whitney}, \binits{D.}},
\bauthor{\bsnm{Phillips}, \binits{E.}},
\bauthor{\bsnm{Chien}, \binits{G.}},
\bauthor{\bsnm{Tompkin}, \binits{J.}},
\bauthor{\bsnm{Konidaris}, \binits{G.}},
\bauthor{\bsnm{Tellex}, \binits{S.}}:
\batitle{Communicating robot arm motion intent through mixed reality head-mounted displays}.
\bjtitle{The International Journal of Robotics Research}
\bvolume{38}(\bissue{12-13}),
\bfpage{1513}--\blpage{1526}
(\byear{2019})
\doiurl{10.1177/0278364919842925}
{\href{https://arxiv.org/abs/https://doi.org/10.1177/0278364919842925}{{https://doi.org/10.1177/0278364919842925}}}
\end{barticle}
\endbibitem

\bibitem[\protect\citeauthoryear{Arevalo~Arboleda et~al.}{2021}]{arevaloarboleda2021assisting}
\begin{bchapter}
\bauthor{\bsnm{Arevalo~Arboleda}, \binits{S.}},
\bauthor{\bsnm{R\"{u}cker}, \binits{F.}},
\bauthor{\bsnm{Dierks}, \binits{T.}},
\bauthor{\bsnm{Gerken}, \binits{J.}}:
\bctitle{Assisting manipulation and grasping in robot teleoperation with augmented reality visual cues}.
In: \bbtitle{{CHI}'21}.
\bpublisher{ACM},
\blocation{New York, NY, USA}
(\byear{2021}).
\doiurl{10.1145/3411764.3445398} .
\burl{https://doi.org/10.1145/3411764.3445398}
\end{bchapter}
\endbibitem

\bibitem[\protect\citeauthoryear{Cogurcu and Maddock}{2023}]{cogurcu2023augmented}
\begin{bchapter}
\bauthor{\bsnm{Cogurcu}, \binits{Y.E.}},
\bauthor{\bsnm{Maddock}, \binits{S.}}:
\bctitle{Augmented reality safety zone configurations in human-robot collaboration: A user study}.
In: \bbtitle{Companion of the 2023 ACM/IEEE International Conference on Human-Robot Interaction}.
\bsertitle{HRI '23},
pp. \bfpage{360}--\blpage{363}.
\bpublisher{ACM},
\blocation{New York, NY, USA}
(\byear{2023}).
\doiurl{10.1145/3568294.3580106} .
\burl{https://doi.org/10.1145/3568294.3580106}
\end{bchapter}
\endbibitem

\bibitem[\protect\citeauthoryear{Gruenefeld et~al.}{2020}]{gruenefeld2020mind}
\begin{bchapter}
\bauthor{\bsnm{Gruenefeld}, \binits{U.}},
\bauthor{\bsnm{Prädel}, \binits{L.}},
\bauthor{\bsnm{Illing}, \binits{J.}},
\bauthor{\bsnm{Stratmann}, \binits{T.}},
\bauthor{\bsnm{Drolshagen}, \binits{S.}},
\bauthor{\bsnm{Pfingsthorn}, \binits{M.}}:
\bctitle{Mind the arm: Realtime visualization of robot motion intent in head-mounted augmented reality}.
In: \bbtitle{Mensch und Computer 2020}
(\byear{2020}).
\doiurl{10.1145/3404983.3405509}
\end{bchapter}
\endbibitem

\bibitem[\protect\citeauthoryear{Andronas et~al.}{2021}]{andronas2021multi}
\begin{barticle}
\bauthor{\bsnm{Andronas}, \binits{D.}},
\bauthor{\bsnm{Apostolopoulos}, \binits{G.}},
\bauthor{\bsnm{Fourtakas}, \binits{N.}},
\bauthor{\bsnm{Makris}, \binits{S.}}:
\batitle{Multi-modal interfaces for natural human-robot interaction}.
\bjtitle{Procedia Manufacturing}
\bvolume{54},
\bfpage{197}--\blpage{202}
(\byear{2021})
\doiurl{10.1016/j.promfg.2021.07.030} .
\bcomment{DET'20}
\end{barticle}
\endbibitem

\bibitem[\protect\citeauthoryear{Rosenholtz et~al.}{2007}]{rosenholtz2007measuring}
\begin{barticle}
\bauthor{\bsnm{Rosenholtz}, \binits{R.}},
\bauthor{\bsnm{Li}, \binits{Y.}},
\bauthor{\bsnm{Nakano}, \binits{L.}}:
\batitle{{Measuring visual clutter}}.
\bjtitle{Journal of Vision}
\bvolume{7}(\bissue{2}),
\bfpage{17}--\blpage{17}
(\byear{2007})
\doiurl{10.1167/7.2.17}
{\href{https://arxiv.org/abs/https://arvojournals.org/arvo/content\_public/journal/jov/932848/jov-7-2-17.pdf}{{https://arvojournals.org/arvo/content\_public/journal/jov/932848/jov-7-2-17.pdf}}}
\end{barticle}
\endbibitem

\bibitem[\protect\citeauthoryear{Lunding et~al.}{2024}]{lunding2024robovisar}
\begin{bchapter}
\bauthor{\bsnm{Lunding}, \binits{R.}},
\bauthor{\bsnm{Lunding}, \binits{M.}},
\bauthor{\bsnm{Feuchtner}, \binits{T.}},
\bauthor{\bsnm{Petersen}, \binits{M.}},
\bauthor{\bsnm{Gr{\o}nb{\ae}k}, \binits{K.}},
\bauthor{\bsnm{Suzuki}, \binits{R.}}:
\bctitle{Robovisar: Immersive authoring of condition-based ar robot visualisations}.
In: \bbtitle{HRI'24}
(\byear{2024})
\end{bchapter}
\endbibitem

\bibitem[\protect\citeauthoryear{Cleaver et~al.}{2021}]{cleaver2021dynamic}
\begin{bchapter}
\bauthor{\bsnm{Cleaver}, \binits{A.}},
\bauthor{\bsnm{Tang}, \binits{D.V.}},
\bauthor{\bsnm{Chen}, \binits{V.}},
\bauthor{\bsnm{Short}, \binits{E.S.}},
\bauthor{\bsnm{Sinapov}, \binits{J.}}:
\bctitle{Dynamic path visualization for human-robot collaboration}.
In: \bbtitle{Companion of the 2021 ACM/IEEE International Conference on Human-Robot Interaction}.
\bsertitle{HRI '21 Companion},
pp. \bfpage{339}--\blpage{343}.
\bpublisher{ACM},
\blocation{New York, NY, USA}
(\byear{2021}).
\doiurl{10.1145/3434074.3447188} .
\burl{https://doi.org/10.1145/3434074.3447188}
\end{bchapter}
\endbibitem

\bibitem[\protect\citeauthoryear{Grubert et~al.}{2023}]{grubert2023text}
\begin{botherref}
\oauthor{\bsnm{Grubert}, \binits{J.}},
\oauthor{\bsnm{Witzani}, \binits{L.}},
\oauthor{\bsnm{Otte}, \binits{A.}},
\oauthor{\bsnm{Gesslein}, \binits{T.}},
\oauthor{\bsnm{Kranz}, \binits{M.}},
\oauthor{\bsnm{Kristensson}, \binits{P.O.}}:
Text {Entry} {Performance} and {Situation} {Awareness} of a {Joint} {Optical} {See}-{Through} {Head}-{Mounted} {Display} and {Smartphone} {System}.
IEEE Transactions on Visualization and Computer Graphics,
1--16
(2023)
\doiurl{10.1109/TVCG.2023.3309316} .
Accessed 2023-08-31
\end{botherref}
\endbibitem

\bibitem[\protect\citeauthoryear{Chan et~al.}{2010}]{chan2010touching}
\begin{botherref}
\oauthor{\bsnm{Chan}, \binits{L.-W.}},
\oauthor{\bsnm{Kao}, \binits{H.-S.}},
\oauthor{\bsnm{Chen}, \binits{M.Y.}},
\oauthor{\bsnm{Lee}, \binits{M.-S.}},
\oauthor{\bsnm{Hsu}, \binits{J.}},
\oauthor{\bsnm{Hung}, \binits{Y.-P.}}:
Touching the {Void}: {Direct}-{Touch} {Interaction} for {Intangible} {Displays}.
CHI'10,
2625--2634
(2010)
\doiurl{10.1145/1753326.1753725}
\end{botherref}
\endbibitem

\bibitem[\protect\citeauthoryear{Elmqvist}{2023}]{elmqvist2023data}
\begin{barticle}
\bauthor{\bsnm{Elmqvist}, \binits{N.}}:
\batitle{Data analytics anywhere and everywhere}.
\bjtitle{Commun. ACM}
\bvolume{66}(\bissue{12}),
\bfpage{52}--\blpage{63}
(\byear{2023})
\doiurl{10.1145/3584858}
\end{barticle}
\endbibitem

\bibitem[\protect\citeauthoryear{Zagermann et~al.}{2022}]{zagermann2022complementary}
\begin{barticle}
\bauthor{\bsnm{Zagermann}, \binits{J.}},
\bauthor{\bsnm{Hubenschmid}, \binits{S.}},
\bauthor{\bsnm{Balestrucci}, \binits{P.}},
\bauthor{\bsnm{Feuchtner}, \binits{T.}},
\bauthor{\bsnm{Mayer}, \binits{S.}},
\bauthor{\bsnm{Ernst}, \binits{M.O.}},
\bauthor{\bsnm{Schmidt}, \binits{A.}},
\bauthor{\bsnm{Reiterer}, \binits{H.}}:
\batitle{Complementary interfaces for visual computing}.
\bjtitle{it - Information Technology}
\bvolume{64}(\bissue{4-5}),
\bfpage{145}--\blpage{154}
(\byear{2022})
\doiurl{10.1515/itit-2022-0031} .
Accessed 2023-08-31
\end{barticle}
\endbibitem

\bibitem[\protect\citeauthoryear{Ens et~al.}{2019}]{ens2019revisiting}
\begin{barticle}
\bauthor{\bsnm{Ens}, \binits{B.}},
\bauthor{\bsnm{Lanir}, \binits{J.}},
\bauthor{\bsnm{Tang}, \binits{A.}},
\bauthor{\bsnm{Bateman}, \binits{S.}},
\bauthor{\bsnm{Lee}, \binits{G.}},
\bauthor{\bsnm{Piumsomboon}, \binits{T.}},
\bauthor{\bsnm{Billinghurst}, \binits{M.}}:
\batitle{Revisiting collaboration through mixed reality: {{The}} evolution of groupware}.
\bjtitle{International Journal of Human-Computer Studies}
\bvolume{131},
\bfpage{81}--\blpage{98}
(\byear{2019})
\doiurl{10.1016/j.ijhcs.2019.05.011}
\end{barticle}
\endbibitem

\bibitem[\protect\citeauthoryear{Sereno et~al.}{2020}]{sereno2020collaborative}
\begin{botherref}
\oauthor{\bsnm{Sereno}, \binits{M.}},
\oauthor{\bsnm{Wang}, \binits{X.}},
\oauthor{\bsnm{Besancon}, \binits{L.}},
\oauthor{\bsnm{Mcguffin}, \binits{M.J.}},
\oauthor{\bsnm{Isenberg}, \binits{T.}}:
Collaborative {{Work}} in {{Augmented Reality}}: {{A Survey}}.
IEEE Transactions on Visualization and Computer Graphics,
1--1
(2020)
\doiurl{10.1109/TVCG.2020.3032761}
\end{botherref}
\endbibitem

\bibitem[\protect\citeauthoryear{Fidalgo et~al.}{2023}]{fidalgo2023survey}
\begin{barticle}
\bauthor{\bsnm{Fidalgo}, \binits{C.G.}},
\bauthor{\bsnm{Yan}, \binits{Y.}},
\bauthor{\bsnm{Cho}, \binits{H.}},
\bauthor{\bsnm{Sousa}, \binits{M.}},
\bauthor{\bsnm{Lindlbauer}, \binits{D.}},
\bauthor{\bsnm{Jorge}, \binits{J.}}:
\batitle{A {{Survey}} on {{Remote Assistance}} and {{Training}} in {{Mixed Reality Environments}}}.
\bjtitle{IEEE TVCG}
\bvolume{29}(\bissue{5}),
\bfpage{2291}--\blpage{2303}
(\byear{2023})
\doiurl{10.1109/TVCG.2023.3247081}
\end{barticle}
\endbibitem

\bibitem[\protect\citeauthoryear{Ratcliffe et~al.}{2021}]{ratcliffe2021extended}
\begin{bchapter}
\bauthor{\bsnm{Ratcliffe}, \binits{J.}},
\bauthor{\bsnm{Soave}, \binits{F.}},
\bauthor{\bsnm{{Bryan-Kinns}}, \binits{N.}},
\bauthor{\bsnm{Tokarchuk}, \binits{L.}},
\bauthor{\bsnm{Farkhatdinov}, \binits{I.}}:
\bctitle{Extended {{Reality}} ({{XR}}) {{Remote Research}}: A {{Survey}} of {{Drawbacks}} and {{Opportunities}}}.
In: \bbtitle{CHI'21},
pp. \bfpage{1}--\blpage{13}.
\bpublisher{ACM},
\blocation{Yokohama Japan}
(\byear{2021}).
\doiurl{10.1145/3411764.3445170}
\end{bchapter}
\endbibitem

\bibitem[\protect\citeauthoryear{Marques et~al.}{2022}]{marques2022conceptual}
\begin{barticle}
\bauthor{\bsnm{Marques}, \binits{B.}},
\bauthor{\bsnm{Silva}, \binits{S.}},
\bauthor{\bsnm{Alves}, \binits{J.}},
\bauthor{\bsnm{Araujo}, \binits{T.}},
\bauthor{\bsnm{Dias}, \binits{P.}},
\bauthor{\bsnm{Santos}, \binits{B.S.}}:
\batitle{A {{Conceptual Model}} and {{Taxonomy}} for {{Collaborative Augmented Reality}}}.
\bjtitle{IEEE Transactions on Visualization and Computer Graphics}
\bvolume{28}(\bissue{12}),
\bfpage{5113}--\blpage{5133}
(\byear{2022})
\doiurl{10.1109/TVCG.2021.3101545}
\end{barticle}
\endbibitem

\bibitem[\protect\citeauthoryear{Marques et~al.}{2024}]{marques2024towards}
\begin{bchapter}
\bauthor{\bsnm{Marques}, \binits{B.}},
\bauthor{\bsnm{Silva}, \binits{S.}},
\bauthor{\bsnm{Pedrosa}, \binits{E.}},
\bauthor{\bsnm{Barros}, \binits{F.}},
\bauthor{\bsnm{Teixeira}, \binits{A.}},
\bauthor{\bsnm{Santos}, \binits{B.S.}}:
\bctitle{Towards unlimited task coverage and direct (far-off) manipulation in extended reality remote collaboration}.
In: \bbtitle{Companion of the 2024 ACM/IEEE International Conference on Human-Robot Interaction}.
\bsertitle{HRI '24},
pp. \bfpage{745}--\blpage{749}.
\bpublisher{Association for Computing Machinery},
\blocation{New York, NY, USA}
(\byear{2024}).
\doiurl{10.1145/3610978.3640737} .
\burl{https://doi.org/10.1145/3610978.3640737}
\end{bchapter}
\endbibitem

\bibitem[\protect\citeauthoryear{Cao et~al.}{2019a}]{cao2019ghostar}
\begin{bchapter}
\bauthor{\bsnm{Cao}, \binits{Y.}},
\bauthor{\bsnm{Wang}, \binits{T.}},
\bauthor{\bsnm{Qian}, \binits{X.}},
\bauthor{\bsnm{Rao}, \binits{P.S.}},
\bauthor{\bsnm{Wadhawan}, \binits{M.}},
\bauthor{\bsnm{Huo}, \binits{K.}},
\bauthor{\bsnm{Ramani}, \binits{K.}}:
\bctitle{Ghostar: A time-space editor for embodied authoring of human-robot collaborative task with augmented reality}.
In: \bbtitle{UIST'19},
pp. \bfpage{521}--\blpage{534}.
\bpublisher{ACM},
\blocation{New York, NY, USA}
(\byear{2019}).
\doiurl{10.1145/3332165.3347902} .
\burl{https://doi.org/10.1145/3332165.3347902}
\end{bchapter}
\endbibitem

\bibitem[\protect\citeauthoryear{Cao et~al.}{2019b}]{cao2019v.ra}
\begin{bchapter}
\bauthor{\bsnm{Cao}, \binits{Y.}},
\bauthor{\bsnm{Xu}, \binits{Z.}},
\bauthor{\bsnm{Li}, \binits{F.}},
\bauthor{\bsnm{Zhong}, \binits{W.}},
\bauthor{\bsnm{Huo}, \binits{K.}},
\bauthor{\bsnm{Ramani}, \binits{K.}}:
\bctitle{V.ra: An in-situ visual authoring system for robot-iot task planning with augmented reality}.
In: \bbtitle{DIS ’19},
pp. \bfpage{1059}--\blpage{1070}.
\bpublisher{ACM},
\blocation{New York, NY, USA}
(\byear{2019}).
\doiurl{10.1145/3322276.3322278} .
\burl{https://doi.org/10.1145/3322276.3322278}
\end{bchapter}
\endbibitem

\bibitem[\protect\citeauthoryear{Fuste et~al.}{2020}]{fuste2020kinetic}
\begin{bchapter}
\bauthor{\bsnm{Fuste}, \binits{A.}},
\bauthor{\bsnm{Reynolds}, \binits{B.}},
\bauthor{\bsnm{Hobin}, \binits{J.}},
\bauthor{\bsnm{Heun}, \binits{V.}}:
\bctitle{Kinetic ar: A framework for robotic motion systems in spatial computing}.
In: \bbtitle{CHI EA '20},
pp. \bfpage{1}--\blpage{8}.
\bpublisher{ACM},
\blocation{New York, NY, USA}
(\byear{2020}).
\doiurl{10.1145/3334480.3382814} .
\burl{https://doi.org/10.1145/3334480.3382814}
\end{bchapter}
\endbibitem

\bibitem[\protect\citeauthoryear{Ikeda and Szafir}{2024}]{ikeda2024programar}
\begin{barticle}
\bauthor{\bsnm{Ikeda}, \binits{B.}},
\bauthor{\bsnm{Szafir}, \binits{D.}}:
\batitle{Programar: Augmented reality end-user robot programming}.
\bjtitle{J. Hum.-Robot Interact.}
(\byear{2024})
\doiurl{10.1145/3640008} .
\bcomment{Just Accepted}
\end{barticle}
\endbibitem

\bibitem[\protect\citeauthoryear{Microsoft}{}]{microsoft}
\begin{botherref}
\oauthor{\bsnm{Microsoft}}:
Introducing Copilot in Microsoft Dynamics 365 Guides, bringing generative AI in mixed reality to frontline workers, \url{https://www.microsoft.com/en-us/dynamics-365/products/guides}.
\url{https://www.microsoft.com/en-us/dynamics-365/products/guides}
\end{botherref}
\endbibitem

\bibitem[\protect\citeauthoryear{Hoang et~al.}{2022}]{hoang2022arviz}
\begin{barticle}
\bauthor{\bsnm{Hoang}, \binits{K.C.}},
\bauthor{\bsnm{Chan}, \binits{W.P.}},
\bauthor{\bsnm{Lay}, \binits{S.}},
\bauthor{\bsnm{Cosgun}, \binits{A.}},
\bauthor{\bsnm{Croft}, \binits{E.A.}}:
\batitle{Arviz: An augmented reality-enabled visualization platform for ros applications}.
\bjtitle{IEEE Robotics and Automation Magazine}
\bvolume{29}(\bissue{1}),
\bfpage{58}--\blpage{67}
(\byear{2022})
\doiurl{10.1109/MRA.2021.3135760}
\end{barticle}
\endbibitem

\bibitem[\protect\citeauthoryear{Ikeda and Szafir}{2022}]{ikeda2022advancing}
\begin{bchapter}
\bauthor{\bsnm{Ikeda}, \binits{B.}},
\bauthor{\bsnm{Szafir}, \binits{D.}}:
\bctitle{Advancing the design of visual debugging tools for roboticists}.
In: \bbtitle{HRI'22},
pp. \bfpage{195}--\blpage{204}.
\bpublisher{IEEE}, \blocation{???}
(\byear{2022}).
\doiurl{10.1109/hri53351.2022.9889392}
\end{bchapter}
\endbibitem

\bibitem[\protect\citeauthoryear{Lee et~al.}{2023}]{lee2023design}
\begin{botherref}
\oauthor{\bsnm{Lee}, \binits{B.}},
\oauthor{\bsnm{Sedlmair}, \binits{M.}},
\oauthor{\bsnm{Schmalstieg}, \binits{D.}}:
Design {{Patterns}} for {{Situated Visualization}} in {{Augmented Reality}}.
IEEE TVCG,
1--12
(2023)
\doiurl{10.1109/TVCG.2023.3327398}
\end{botherref}
\endbibitem

\bibitem[\protect\citeauthoryear{Martins et~al.}{2022}]{martins2022augmented}
\begin{barticle}
\bauthor{\bsnm{Martins}, \binits{N.C.}},
\bauthor{\bsnm{Marques}, \binits{B.}},
\bauthor{\bsnm{Alves}, \binits{J.}},
\bauthor{\bsnm{Ara{\'u}jo}, \binits{T.}},
\bauthor{\bsnm{Dias}, \binits{P.}},
\bauthor{\bsnm{Santos}, \binits{B.S.}}:
\batitle{Augmented reality situated visualization in decision-making}.
\bjtitle{Multimedia Tools and Applications}
\bvolume{81}(\bissue{11}),
\bfpage{14749}--\blpage{14772}
(\byear{2022})
\doiurl{10.1007/s11042-021-10971-4}
\end{barticle}
\endbibitem

\bibitem[\protect\citeauthoryear{Leiva et~al.}{2021}]{leiva2021rapido}
\begin{bbook}
\bauthor{\bsnm{Leiva}, \binits{G.}},
\bauthor{\bsnm{Gr\o{}nb\ae{}k}, \binits{J.E.}},
\bauthor{\bsnm{Klokmose}, \binits{C.N.}},
\bauthor{\bsnm{Nguyen}, \binits{C.}},
\bauthor{\bsnm{Kazi}, \binits{R.H.}},
\bauthor{\bsnm{Asente}, \binits{P.}}:
\bbtitle{Rapido: Prototyping Interactive AR Experiences through Programming by Demonstration},
pp. \bfpage{626}--\blpage{637}.
\bpublisher{ACM},
\blocation{New York, NY, USA}
(\byear{2021}).
\burl{https://doi.org/10.1145/3472749.3474774}
\end{bbook}
\endbibitem

\bibitem[\protect\citeauthoryear{Ghamandi et~al.}{2023}]{ghamandi2023what}
\begin{bchapter}
\bauthor{\bsnm{Ghamandi}, \binits{R.K.}},
\bauthor{\bsnm{Hmaiti}, \binits{Y.}},
\bauthor{\bsnm{Nguyen}, \binits{T.T.}},
\bauthor{\bsnm{Ghasemaghaei}, \binits{A.}},
\bauthor{\bsnm{Kattoju}, \binits{R.K.}},
\bauthor{\bsnm{Taranta}, \binits{E.M.}},
\bauthor{\bsnm{LaViola}, \binits{J.J.}}:
\bctitle{What {{And How Together}}: {{A Taxonomy On}} 30 {{Years Of Collaborative Human-Centered XR Tasks}}}.
In: \bbtitle{{{ISMAR}}},
pp. \bfpage{322}--\blpage{335}.
\bpublisher{IEEE},
\blocation{Sydney, Australia}
(\byear{2023}).
\doiurl{10.1109/ISMAR59233.2023.00047}
\end{bchapter}
\endbibitem

\bibitem[\protect\citeauthoryear{Walker et~al.}{2023}]{walker2023virtual}
\begin{botherref}
\oauthor{\bsnm{Walker}, \binits{M.}},
\oauthor{\bsnm{Phung}, \binits{T.}},
\oauthor{\bsnm{Chakraborti}, \binits{T.}},
\oauthor{\bsnm{Williams}, \binits{T.}},
\oauthor{\bsnm{Szafir}, \binits{D.}}:
Virtual, augmented, and mixed reality for human-robot interaction: A survey and virtual design element taxonomy.
J. Hum.-Robot Interact.
\textbf{12}(4)
(2023)
\doiurl{10.1145/3597623}
\end{botherref}
\endbibitem

\bibitem[\protect\citeauthoryear{Chan et~al.}{2020}]{chan2020augmented}
\begin{bchapter}
\bauthor{\bsnm{Chan}, \binits{W.P.}},
\bauthor{\bsnm{Hanks}, \binits{G.}},
\bauthor{\bsnm{Sakr}, \binits{M.}},
\bauthor{\bsnm{Zuo}, \binits{T.}},
\bauthor{\bsnm{Loos}, \binits{H.F.}},
\bauthor{\bsnm{Croft}, \binits{E.}}:
\bctitle{An augmented reality human-robot physical collaboration interface design for shared, large-scale, labour-intensive manufacturing tasks}.
In: \bbtitle{2020 IEEE/RSJ International Conference on Intelligent Robots and Systems (IROS)},
pp. \bfpage{11308}--\blpage{11313}.
\bpublisher{IEEE Press},
\blocation{Las Vegas, NV, USA}
(\byear{2020}).
\doiurl{10.1109/IROS45743.2020.9341119} .
\burl{https://ieeexplore.ieee.org/document/9341119}
\end{bchapter}
\endbibitem

\bibitem[\protect\citeauthoryear{Andersen et~al.}{2016}]{andersen2016projecting}
\begin{bchapter}
\bauthor{\bsnm{Andersen}, \binits{R.S.}},
\bauthor{\bsnm{Madsen}, \binits{O.}},
\bauthor{\bsnm{Moeslund}, \binits{T.B.}},
\bauthor{\bsnm{Amor}, \binits{H.B.}}:
\bctitle{Projecting robot intentions into human environments}.
In: \bbtitle{2016 25th IEEE International Symposium on Robot and Human Interactive Communication (RO-MAN)},
pp. \bfpage{294}--\blpage{301}.
\bpublisher{Institute of Electrical and Electronics Engineers Inc.},
\blocation{United States}
(\byear{2016}).
\doiurl{10.1109/ROMAN.2016.7745145}
\end{bchapter}
\endbibitem

\bibitem[\protect\citeauthoryear{Li et~al.}{2019}]{li2019research}
\begin{barticle}
\bauthor{\bsnm{Li}, \binits{W.}},
\bauthor{\bsnm{Wang}, \binits{J.}},
\bauthor{\bsnm{Jiao}, \binits{S.}},
\bauthor{\bsnm{Wang}, \binits{M.}},
\bauthor{\bsnm{Li}, \binits{S.}}:
\batitle{Research on the visual elements of augmented reality assembly processes}.
\bjtitle{Virtual Reality and Intelligent Hardware}
\bvolume{1}(\bissue{6}),
\bfpage{622}--\blpage{634}
(\byear{2019})
\doiurl{10.1016/j.vrih.2019.09.006} .
\bcomment{VR/AR in industry}
\end{barticle}
\endbibitem

\bibitem[\protect\citeauthoryear{Brudy et~al.}{2019}]{brudy2019crossdevice}
\begin{bchapter}
\bauthor{\bsnm{Brudy}, \binits{F.}},
\bauthor{\bsnm{Holz}, \binits{C.}},
\bauthor{\bsnm{Rädle}, \binits{R.}},
\bauthor{\bsnm{Wu}, \binits{C.-J.}},
\bauthor{\bsnm{Houben}, \binits{S.}},
\bauthor{\bsnm{Klokmose}, \binits{C.N.}},
\bauthor{\bsnm{Marquardt}, \binits{N.}}:
\bctitle{Cross-{Device} {Taxonomy}: {Survey}, {Opportunities} and {Challenges} of {Interactions} {Spanning} {Across} {Multiple} {Devices}}.
In: \bbtitle{{CHI}'19},
pp. \bfpage{1}--\blpage{28}.
\bpublisher{ACM Press},
\blocation{Glasgow, Scotland Uk}
(\byear{2019}).
\doiurl{10.1145/3290605.3300792} .
\burl{http://dl.acm.org/citation.cfm?doid=3290605.3300792}
Accessed 2019-11-10
\end{bchapter}
\endbibitem

\bibitem[\protect\citeauthoryear{Feiner and Shamash}{1991}]{feiner1991hybrid}
\begin{bchapter}
\bauthor{\bsnm{Feiner}, \binits{S.}},
\bauthor{\bsnm{Shamash}, \binits{A.}}:
\bctitle{Hybrid user interfaces: breeding virtually bigger interfaces for physically smaller computers}.
In: \bbtitle{Proceedings of the 4th Annual {ACM} Symposium on {User} Interface Software and Technology}.
\bsertitle{{UIST} '91},
pp. \bfpage{9}--\blpage{17}.
\bpublisher{ACM},
\blocation{Hilton Head, South Carolina, USA}
(\byear{1991}).
\doiurl{10.1145/120782.120783} .
\burl{https://doi.org/10.1145/120782.120783}
Accessed 2020-04-29
\end{bchapter}
\endbibitem

\bibitem[\protect\citeauthoryear{Hubenschmid et~al.}{2023}]{hubenschmid2023smartphone}
\begin{bchapter}
\bauthor{\bsnm{Hubenschmid}, \binits{S.}},
\bauthor{\bsnm{Zagermann}, \binits{J.}},
\bauthor{\bsnm{Leicht}, \binits{D.}},
\bauthor{\bsnm{Reiterer}, \binits{H.}},
\bauthor{\bsnm{Feuchtner}, \binits{T.}}:
\bctitle{{ARound} the {Smartphone}: {Investigating} the {Effects} of {Virtually}-{Extended} {Display} {Size} on {Spatial} {Memory}}.
In: \bbtitle{CHI '23},
pp. \bfpage{1}--\blpage{15}.
\bpublisher{ACM},
\blocation{Hamburg Germany}
(\byear{2023}).
\doiurl{10.1145/3544548.3581438} .
\burl{https://dl.acm.org/doi/10.1145/3544548.3581438}
Accessed 2023-04-22
\end{bchapter}
\endbibitem

\bibitem[\protect\citeauthoryear{Langner et~al.}{2021}]{langner2021marvis}
\begin{bchapter}
\bauthor{\bsnm{Langner}, \binits{R.}},
\bauthor{\bsnm{Satkowski}, \binits{M.}},
\bauthor{\bsnm{Büschel}, \binits{W.}},
\bauthor{\bsnm{Dachselt}, \binits{R.}}:
\bctitle{{MARVIS}: {Combining} {Mobile} {Devices} and {Augmented} {Reality} for {Visual} {Data} {Analysis}}.
In: \bbtitle{{CHI}'21}.
\bsertitle{{CHI} '21},
pp. \bfpage{1}--\blpage{17}.
\bpublisher{ACM},
\blocation{New York, NY, USA}
(\byear{2021}).
\doiurl{10.1145/3411764.3445593} .
\burl{https://doi.org/10.1145/3411764.3445593}
Accessed 2021-05-10
\end{bchapter}
\endbibitem

\bibitem[\protect\citeauthoryear{Reipschläger et~al.}{2021}]{reipschlager2021personal}
\begin{barticle}
\bauthor{\bsnm{Reipschläger}, \binits{P.}},
\bauthor{\bsnm{Flemisch}, \binits{T.}},
\bauthor{\bsnm{Dachselt}, \binits{R.}}:
\batitle{Personal {Augmented} {Reality} for {Information} {Visualization} on {Large} {Interactive} {Displays}}.
\bjtitle{IEEE TVCG}
(\byear{2021})
\doiurl{10.1109/TVCG.2020.3030460} .
Accessed 2020-09-09
\end{barticle}
\endbibitem

\bibitem[\protect\citeauthoryear{Büschel et~al.}{2019}]{buschel2019investigating}
\begin{bchapter}
\bauthor{\bsnm{Büschel}, \binits{W.}},
\bauthor{\bsnm{Mitschick}, \binits{A.}},
\bauthor{\bsnm{Meyer}, \binits{T.}},
\bauthor{\bsnm{Dachselt}, \binits{R.}}:
\bctitle{Investigating {Smartphone}-based {Pan} and {Zoom} in {3D} {Data} {Spaces} in {Augmented} {Reality}}.
In: \bbtitle{{MobileHCI} '19},
pp. \bfpage{1}--\blpage{13}.
\bpublisher{ACM Press},
\blocation{Taipei, Taiwan}
(\byear{2019}).
\doiurl{10.1145/3338286.3340113} .
\burl{http://dl.acm.org/citation.cfm?doid=3338286.3340113}
Accessed 2019-11-16
\end{bchapter}
\endbibitem

\bibitem[\protect\citeauthoryear{Butscher et~al.}{2018}]{butscher2018clusters}
\begin{bchapter}
\bauthor{\bsnm{Butscher}, \binits{S.}},
\bauthor{\bsnm{Hubenschmid}, \binits{S.}},
\bauthor{\bsnm{Müller}, \binits{J.}},
\bauthor{\bsnm{Fuchs}, \binits{J.}},
\bauthor{\bsnm{Reiterer}, \binits{H.}}:
\bctitle{Clusters, {Trends}, and {Outliers}: {How} {Immersive} {Technologies} {Can} {Facilitate} the {Collaborative} {Analysis} of {Multidimensional} {Data}}.
In: \bbtitle{{CHI}'18},
pp. \bfpage{1}--\blpage{12}.
\bpublisher{ACM Press},
\blocation{New York, New York, USA}
(\byear{2018}).
\doiurl{10.1145/3173574.3173664} .
\burl{http://dl.acm.org/citation.cfm?doid=3173574.3173664}
\end{bchapter}
\endbibitem

\bibitem[\protect\citeauthoryear{Hubenschmid et~al.}{2021}]{hubenschmid2021stream}
\begin{bchapter}
\bauthor{\bsnm{Hubenschmid}, \binits{S.}},
\bauthor{\bsnm{Zagermann}, \binits{J.}},
\bauthor{\bsnm{Butscher}, \binits{S.}},
\bauthor{\bsnm{Reiterer}, \binits{H.}}:
\bctitle{{STREAM}: {Exploring} the {Combination} of {Spatially}-{Aware} {Tablets} with {Augmented} {Reality} {Head}-{Mounted} {Displays} for {Immersive} {Analytics}}.
In: \bbtitle{CHI '21},
pp. \bfpage{1}--\blpage{14}.
\bpublisher{ACM},
\blocation{Yokohama Japan}
(\byear{2021}).
\doiurl{10.1145/3411764.3445298} .
\burl{https://doi.org/10.1145/3411764.3445298}
Accessed 2021-05-10
\end{bchapter}
\endbibitem

\bibitem[\protect\citeauthoryear{Knierim et~al.}{2021}]{knierim2021smartphone}
\begin{barticle}
\bauthor{\bsnm{Knierim}, \binits{P.}},
\bauthor{\bsnm{Hein}, \binits{D.}},
\bauthor{\bsnm{Schmidt}, \binits{A.}},
\bauthor{\bsnm{Kosch}, \binits{T.}}:
\batitle{The {SmARtphone} {Controller}: {Leveraging} {Smartphones} as {Input} and {Output} {Modality} for {Improved} {Interaction} within {Mobile} {Augmented} {Reality} {Environments}}.
\bjtitle{i-com}
\bvolume{20}(\bissue{1}),
\bfpage{49}--\blpage{61}
(\byear{2021})
\doiurl{10.1515/icom-2021-0003} .
Accessed 2021-06-15
\end{barticle}
\endbibitem

\bibitem[\protect\citeauthoryear{Hubenschmid et~al.}{2021}]{hubenschmid2021asynchronous}
\begin{bchapter}
\bauthor{\bsnm{Hubenschmid}, \binits{S.}},
\bauthor{\bsnm{Zagermann}, \binits{J.}},
\bauthor{\bsnm{Fink}, \binits{D.}},
\bauthor{\bsnm{Wieland}, \binits{J.}},
\bauthor{\bsnm{Feuchtner}, \binits{T.}},
\bauthor{\bsnm{Reiterer}, \binits{H.}}:
\bctitle{Towards asynchronous hybrid user interfaces for cross-reality interaction}.
In: \bbtitle{{ISS}'21 Workshop Proceedings: "{Transitional} {Interfaces} in {Mixed} and {Cross}-{Reality}: {A} new frontier?"}
(\byear{2021}).
\doiurl{10.18148/kops/352-2-84mm0sggczq02}
\end{bchapter}
\endbibitem

\bibitem[\protect\citeauthoryear{Hubenschmid et~al.}{2022}]{hubenschmid2022relive}
\begin{bchapter}
\bauthor{\bsnm{Hubenschmid}, \binits{S.}},
\bauthor{\bsnm{Wieland}, \binits{J.}},
\bauthor{\bsnm{Fink}, \binits{D.I.}},
\bauthor{\bsnm{Batch}, \binits{A.}},
\bauthor{\bsnm{Zagermann}, \binits{J.}},
\bauthor{\bsnm{Elmqvist}, \binits{N.}},
\bauthor{\bsnm{Reiterer}, \binits{H.}}:
\bctitle{{ReLive}: {Bridging} {In}-{Situ} and {Ex}-{Situ} {Visual} {Analytics} for {Analyzing} {Mixed} {Reality} {User} {Studies}}.
In: \bbtitle{CHI'22},
pp. \bfpage{1}--\blpage{20}.
\bpublisher{ACM},
\blocation{New Orleans LA USA}
(\byear{2022}).
\doiurl{10.1145/3491102.3517550} .
\burl{https://dl.acm.org/doi/10.1145/3491102.3517550}
Accessed 2022-05-01
\end{bchapter}
\endbibitem

\bibitem[\protect\citeauthoryear{Kim and Dey}{2009}]{kim2009simulated}
\begin{bchapter}
\bauthor{\bsnm{Kim}, \binits{S.}},
\bauthor{\bsnm{Dey}, \binits{A.K.}}:
\bctitle{Simulated augmented reality windshield display as a cognitive mapping aid for elder driver navigation}.
In: \bbtitle{Proceedings of the {SIGCHI} {Conference} on {Human} {Factors} in {Computing} {Systems}},
pp. \bfpage{133}--\blpage{142}.
\bpublisher{ACM},
\blocation{Boston MA USA}
(\byear{2009}).
\doiurl{10.1145/1518701.1518724} .
\burl{https://dl.acm.org/doi/10.1145/1518701.1518724}
Accessed 2024-03-04
\end{bchapter}
\endbibitem

\bibitem[\protect\citeauthoryear{Rogers and Monsell}{1995}]{rogers1995costs}
\begin{barticle}
\bauthor{\bsnm{Rogers}, \binits{R.D.}},
\bauthor{\bsnm{Monsell}, \binits{S.}}:
\batitle{Costs of a predictible switch between simple cognitive tasks.}
\bjtitle{Journal of Experimental Psychology: General}
\bvolume{124}(\bissue{2}),
\bfpage{207}--\blpage{231}
(\byear{1995})
\doiurl{10.1037/0096-3445.124.2.207} .
Accessed 2024-03-04
\end{barticle}
\endbibitem

\bibitem[\protect\citeauthoryear{Rashid et~al.}{2012a}]{rashid2012cost}
\begin{bchapter}
\bauthor{\bsnm{Rashid}, \binits{U.}},
\bauthor{\bsnm{Nacenta}, \binits{M.A.}},
\bauthor{\bsnm{Quigley}, \binits{A.}}:
\bctitle{The cost of display switching: a comparison of mobile, large display and hybrid {UI} configurations}.
In: \bbtitle{AVI'12},
pp. \bfpage{99}--\blpage{106}.
\bpublisher{ACM},
\blocation{Capri Island, Italy}
(\byear{2012}).
\burl{https://doi.org/10.1145/2254556.2254577}
Accessed 2020-03-19
\end{bchapter}
\endbibitem

\bibitem[\protect\citeauthoryear{Rashid et~al.}{2012b}]{rashid2012factors}
\begin{bchapter}
\bauthor{\bsnm{Rashid}, \binits{U.}},
\bauthor{\bsnm{Nacenta}, \binits{M.A.}},
\bauthor{\bsnm{Quigley}, \binits{A.}}:
\bctitle{Factors influencing visual attention switch in multi-display user interfaces: a survey}.
In: \bbtitle{Proceedings of the 2012 {International} {Symposium} on {Pervasive} {Displays} - {PerDis} '12},
pp. \bfpage{1}--\blpage{6}.
\bpublisher{ACM Press},
\blocation{Porto, Portugal}
(\byear{2012}).
\doiurl{10.1145/2307798.2307799} .
\burl{http://dl.acm.org/citation.cfm?doid=2307798.2307799}
Accessed 2022-05-01
\end{bchapter}
\endbibitem

\bibitem[\protect\citeauthoryear{\relax Universal~Robots}{}]{ur5e}
\begin{botherref}
\oauthor{\bsnm{Universal~Robots}}:
Robot arm Technical specification \url{https://www.universal-robots.com/media/1826690/01_2023_collective_data_sheet-1.pdf}, (May 09, 2024).
\url{https://www.universal-robots.com/media/1826690/01_2023_collective_data_sheet-1.pdf}
\end{botherref}
\endbibitem

\bibitem[\protect\citeauthoryear{Park et~al.}{2023}]{park2023digitalizing}
\begin{bchapter}
\bauthor{\bsnm{Park}, \binits{J.}},
\bauthor{\bsnm{Casser}, \binits{F.}},
\bauthor{\bsnm{Schlette}, \binits{C.}}:
\bctitle{Digitalizing manual processes using digital twins and product lifecycle management for safe human-robot interaction scenarios}.
In: \bbtitle{2023 28th International Conference on Automation and Computing (ICAC)},
pp. \bfpage{1}--\blpage{6}
(\byear{2023}).
\doiurl{10.1109/ICAC57885.2023.10275241}
\end{bchapter}
\endbibitem

\bibitem[\protect\citeauthoryear{{\relax OASIS MQTT Technical Committee}}{}]{mqtt}
\begin{botherref}
\oauthor{\bsnm{{\relax OASIS MQTT Technical Committee}}}:
{MQTT}, \url{https://mqtt.org}, (May 09, 2024).
\url{https://mqtt.org}
\end{botherref}
\endbibitem

\bibitem[\protect\citeauthoryear{S{\"a}{\"a}ski et~al.}{2008}]{saeaeski2008integration}
\begin{bchapter}
\bauthor{\bsnm{S{\"a}{\"a}ski}, \binits{J.}},
\bauthor{\bsnm{Salonen}, \binits{T.}},
\bauthor{\bsnm{Hakkarainen}, \binits{M.}},
\bauthor{\bsnm{Siltanen}, \binits{S.}},
\bauthor{\bsnm{Woodward}, \binits{C.}},
\bauthor{\bsnm{Lempi{\"a}inen}, \binits{J.}}:
\bctitle{Integration of design and assembly using augmented reality}.
In: \bbtitle{Micro-Assembly Technologies and Applications},
pp. \bfpage{395}--\blpage{404}.
\bpublisher{Springer},
\blocation{Boston, MA}
(\byear{2008}).
\doiurl{10.1007/978-0-387-77405-3_39}
\end{bchapter}
\endbibitem

\bibitem[\protect\citeauthoryear{{\relax Universal Robots}}{}]{rtde}
\begin{botherref}
\oauthor{\bsnm{{\relax Universal Robots}}}:
RTDE client library - Python \url{https://github.com/UniversalRobots/RTDE_Python_Client_Library}, (Feb 22, 2024).
\url{https://github.com/UniversalRobots/RTDE_Python_Client_Library}
\end{botherref}
\endbibitem

\bibitem[\protect\citeauthoryear{{\relax Eclipse foundation}}{}]{mosquitto}
\begin{botherref}
\oauthor{\bsnm{{\relax Eclipse foundation}}}:
Mosquitto, \url{https://mosquitto.org}, (May 09, 2024).
\url{https://mosquitto.org}
\end{botherref}
\endbibitem

\bibitem[\protect\citeauthoryear{Ledo et~al.}{2018}]{ledo2018evaluation}
\begin{bchapter}
\bauthor{\bsnm{Ledo}, \binits{D.}},
\bauthor{\bsnm{Houben}, \binits{S.}},
\bauthor{\bsnm{Vermeulen}, \binits{J.}},
\bauthor{\bsnm{Marquardt}, \binits{N.}},
\bauthor{\bsnm{Oehlberg}, \binits{L.}},
\bauthor{\bsnm{Greenberg}, \binits{S.}}:
\bctitle{Evaluation strategies for hci toolkit research}.
In: \bbtitle{CHI'18},
pp. \bfpage{1}--\blpage{17}.
\bpublisher{ACM},
\blocation{New York, NY, USA}
(\byear{2018}).
\doiurl{10.1145/3173574.3173610} .
\burl{https://doi.org/10.1145/3173574.3173610}
\end{bchapter}
\endbibitem

\bibitem[\protect\citeauthoryear{De~Franco et~al.}{2019}]{defranco2019intuitive}
\begin{bchapter}
\bauthor{\bsnm{De~Franco}, \binits{A.}},
\bauthor{\bsnm{Lamon}, \binits{E.}},
\bauthor{\bsnm{Balatti}, \binits{P.}},
\bauthor{\bsnm{De~Momi}, \binits{E.}},
\bauthor{\bsnm{Ajoudani}, \binits{A.}}:
\bctitle{An intuitive augmented reality interface for task scheduling, monitoring, and work performance improvement in human-robot collaboration}.
In: \bbtitle{2019 IEEE International Work Conference on Bioinspired Intelligence (IWOBI)},
pp. \bfpage{75}--\blpage{80}
(\byear{2019}).
\doiurl{10.1109/IWOBI47054.2019.9114472}
\end{bchapter}
\endbibitem

\bibitem[\protect\citeauthoryear{Renner et~al.}{2018}]{renner2018wysiwicd}
\begin{bchapter}
\bauthor{\bsnm{Renner}, \binits{P.}},
\bauthor{\bsnm{Lier}, \binits{F.}},
\bauthor{\bsnm{Friese}, \binits{F.}},
\bauthor{\bsnm{Pfeiffer}, \binits{T.}},
\bauthor{\bsnm{Wachsmuth}, \binits{S.}}:
\bctitle{Wysiwicd: What you see is what i can do}.
In: \bbtitle{Companion of the 2018 ACM/IEEE International Conference on Human-Robot Interaction}.
\bsertitle{HRI '18},
p. \bfpage{382}.
\bpublisher{ACM},
\blocation{New York, NY, USA}
(\byear{2018}).
\doiurl{10.1145/3173386.3177533} .
\burl{https://doi.org/10.1145/3173386.3177533}
\end{bchapter}
\endbibitem

\end{thebibliography}

\end{document}